\documentclass{LMCS}

\def\dOi{11(3:9)2015}
\lmcsheading%
{\dOi}
{1--28}
{}
{}
{Apr.~25, 2014}
{Sep.~11, 2015}
{}

\ACMCCS{[{\bf Theory of computation}]: Logic---Finite Model Theory}

\usepackage{amsmath}
\usepackage{amssymb}
\usepackage{mathrsfs}  
\usepackage{hyperref}
\usepackage{graphicx}
\usepackage{color}
\usepackage[all]{xy}
\usepackage{subfig}

\newtheorem{definition}{Definition}
\newtheorem{ex}{Example}
\newtheorem*{remarks}{Remarks}
\newtheorem*{remark}{Remark}

\newcommand{\csplogic}{\ensuremath{\{\exists, \wedge \}
    \mbox{-}\mathrm{FO}}}
\newcommand{\qcsplogic}{\ensuremath{\{\exists, \forall, \wedge \}
    \mbox{-}\mathrm{FO}}}
\newcommand{\qcsplogiceq}{\ensuremath{\{\exists, \forall, \wedge,= \} \mbox{-}\mathrm{FO}}}
\newcommand{\mylogic}{\ensuremath{\{\exists, \forall, \wedge,\vee \} \mbox{-}\mathrm{FO}}}
\newcommand{\posFO}{\ensuremath{\{\exists, \forall, \wedge,\vee,= \}
    \mbox{-}\mathrm{FO}}}
\newcommand{\tuple}[1]{\ensuremath{\mathbf{#1}}}

\newcommand{\notmodels}{\ensuremath{ \models \hspace{-3mm} / \hspace{2mm} }}
\newcommand{\rank}[0]{\ensuremath{\mathsf{rank}}}
\newcommand{\depth}[0]{\ensuremath{\mathsf{depth}}}

\newcommand{\skolem}[0]{\ensuremath{\mathsf{Skolem}}}
\newcommand{\surhom}{
  \ensuremath{
      \negthinspace 
      \longrightarrow
      \hspace{-5mm} \rightarrow \hspace{1mm}
  }
}
\newcommand{\homm}{
  \ensuremath{
      \negthinspace 
      \longrightarrow
      \negthinspace
  }
}
\newcommand{\nosurhom}{
  \ensuremath{
      \negthinspace 
      \longrightarrow
      \hspace{-5mm} \rightarrow \hspace{-4mm} / \hspace{3mm}
  }
}
\renewcommand{\phi}{\varphi}

\title[Quantified Constraints and Containment Problems]
        {Quantified Constraints and Containment Problems\rsuper*}

\author[H.~Chen]{Hubie Chen\rsuper a}
\address{{\lsuper a}Departamento LSI,
Facultad de Inform\'{a}tica,
Universidad del Pa\'{i}s Vasco,
E-20018 San Sebasti\'{a}n,
Spain
and
IKERBASQUE, Basque Foundation for Science,
E-48011 Bilbao,
Spain}
\email{hubie.chen@ehu.es}

\author[F.~Madelaine]{Florent Madelaine\rsuper b}
\address{{\lsuper b}Clermont Universit{\'e}, Universit{\'e} d'Auvergne, 
LIMOS, BP 10448, F-63000 Clermont-Ferrand, France}
\email{florent.madelaine@udamail.fr}

\author[B.~Martin]{Barnaby Martin\rsuper c}
\address{{\lsuper c}Science and Technology, Middlesex University,
The Burroughs, Hendon, London NW4 4BT, U.K.}
\email{barnabymartin@gmail.com}

\thanks{{\lsuper c}The third author was supported by EPSRC grant EP/L005654/1.}
\titlecomment{{\lsuper*}The main result of this paper has appeared in the extended abstracts \cite{LICS2008} and \cite{QCores}.}

\begin{document}

\maketitle

\begin{abstract}
The quantified constraint satisfaction problem $\mathrm{QCSP}(\mathcal{A})$ is the problem to decide whether a positive Horn sentence, involving nothing more than the two quantifiers and conjunction, is true on some fixed structure $\mathcal{A}$. We study two containment problems related to the QCSP.

Firstly, we give a combinatorial condition on finite structures $\mathcal{A}$ and $\mathcal{B}$ that is necessary and sufficient to render $\mathrm{QCSP}(\mathcal{A}) \subseteq \mathrm{QCSP}(\mathcal{B})$.  We prove that $\mathrm{QCSP}(\mathcal{A}) \subseteq \mathrm{QCSP}(\mathcal{B})$, that is all sentences of positive Horn logic true on $\mathcal{A}$ are true on $\mathcal{B}$, iff there is a surjective homomorphism from $\mathcal{A}^{|A|^{|B|}}$ to $\mathcal{B}$. This can be seen as improving an old result of Keisler that shows the former equivalent to there being a surjective homomorphism from $\mathcal{A}^\omega$ to $\mathcal{B}$.  We note that this condition is already necessary to guarantee containment of the $\Pi_2$ restriction of the QCSP, that is $\Pi_2$-$\mathrm{CSP}(\mathcal{A}) \subseteq \Pi_2$-$\mathrm{CSP}(\mathcal{B})$. 
The exponent's bound of ${|A|^{|B|}}$ places the decision procedure for the model containment problem in non-deterministic double-exponential time complexity. We further show the exponent's bound $|A|^{|B|}$ to be close to tight by giving a sequence of structures $\mathcal{A}$ together with a fixed $\mathcal{B}$, $|B|=2$, such that there is a surjective homomorphism from $\mathcal{A}^r$ to $\mathcal{B}$ only when $r \geq |A|$.

Secondly, we prove that the entailment problem for positive Horn fragment of first-order logic is decidable. That is, given two sentences $\varphi$ and $\psi$ of positive Horn,  we give an algorithm that determines whether $\varphi \rightarrow \psi$ is true in all structures (models). Our result is in some sense tight, since we show that the entailment problem for positive first-order logic (\mbox{i.e.} positive Horn plus disjunction) is undecidable.

In the final part of the paper we ponder a notion of Q-core that is some canonical representative among the class of templates that engender the same QCSP. Although the Q-core is not as well-behaved as its better known cousin the core, we demonstrate that it is still a useful notion in the realm of QCSP complexity classifications.
\end{abstract}

\section{Introduction}

The \emph{constraint satisfaction problem} (CSP), much studied in artificial intelligence, is known to admit several equivalent formulations, two of the most popular of which are the model-checking problem for primitive positive first-order sentences and the homomorphism problem (see, e.g., \cite{KolaitisVardiBook05}). The CSP is NP-complete in general, and a great deal of effort has been expended in classifying its complexity for certain restricted cases, in particular where it is parameterised by the \emph{constraint language} (which corresponds to the model in the model-checking problem and the right-hand structure of the homomorphism problem). The problems $\mathrm{CSP}(\mathcal{A})$ thereby obtained, sometimes termed non-uniform \cite{FederVardi}, are conjectured \cite{FederVardi,Bulatov00:algebras} to be always polynomial-time tractable or NP-complete. While this has not been settled in general, a number of partial results are known (e.g. over structures of size $\leq 3$ \cite{Schaefer,BulatovJACM} and over smooth digraphs graphs \cite{HellNesetril,barto:1782}). Most of the great advances in these complexity classifications in the past decade have been driven by the algebraic method (\mbox{e.g.} \cite{BulatovJACM,barto:1782}). This involves studying indirectly the relations of a structure through certain operations called polymorphisms that preserve them.

The model containment problem for CSP is the question, for finite structures $\mathcal{A}$ and $\mathcal{B}$, whether $\mathrm{CSP}(\mathcal{A}) \subseteq \mathrm{CSP}(\mathcal{B})$? It is easy to see that this is equivalent to the question of existence of a homomorphism from $\mathcal{A}$ to $\mathcal{B}$. Thus the model containment problem for CSP is, essentially, a CSP itself. The condition for $\mathrm{CSP}(\mathcal{A}) = \mathrm{CSP}(\mathcal{B})$ is, therefore, that $\mathcal{A}$ and $\mathcal{B}$ are homomorphically equivalent. This in turn is equivalent to the condition that $\mathcal{A}$ and $\mathcal{B}$ share the same, or rather isomorphic, \emph{cores} (where the core of a structure $\mathcal{A}$ is a minimal substructure that is homomorphically equivalent to $\mathcal{A}$). The complexity classification problem for $\mathrm{CSP}(\mathcal{A})$ is greatly facilitated by the fact that we may, therefore, assume that $\mathcal{A}$ is a core -- i.e. that $\mathcal{A}$ is a minimal representative of its equivalence class under the equivalence relation induced by homomorphic equivalence.

A useful generalisation of the CSP involves considering the model-checking problem for positive Horn (pH)  sentences (where we add to primitive positive logic universal quantification). This allows for a broader class of problems, used in artificial intelligence to capture non-monotonic reasoning, whose complexities rise through the polynomial hierarchy up to Pspace. 
When the quantifier prefix is restricted to $\Pi_2$, with all universal quantifiers preceding existential quantifiers, we obtain the $\Pi_2$-CSP; when the prefix is unrestricted, we obtain the \emph{quantified constraint satisfaction problem} (QCSP). In general, the $\Pi_2$-CSP and QCSP  are $\Pi^{\mathrm{P}}_{2}$-complete and Pspace-complete, respectively (for more on these complexity classes, we direct the reader to \cite{ComputationalComplexity}).
As with the CSP, it has become popular to consider the QCSP parameterised by the constraint language, \mbox{i.e.} the model in the model-checking problem, and there is a growing body of results delineating the tractable instances from those that are (probably) intractable \cite{OxfordQuantifiedConstraints,chen-2006}. 

The model containment problem for QCSP takes as input two finite structures $\mathcal{A}$ and $\mathcal{B}$ and asks whether $\mathrm{QCSP}(\mathcal{A}) \subseteq \mathrm{QCSP}(\mathcal{B})$. Unlike the situation with the CSP, it is not apparent that this containment problem is in any way similar to the QCSP itself. As far as we know, neither a characterisation nor an algorithm for this problem had been known. In this paper we provide both, \mbox{i.e.} we settle the question as to when exactly $\mathrm{QCSP}(\mathcal{A}) \subseteq \mathrm{QCSP}(\mathcal{B})$ by giving a characterising morphism from $\mathcal{A}$ to $\mathcal{B}$.  It turns out that $\mathrm{QCSP}(\mathcal{A}) \subseteq \mathrm{QCSP}(\mathcal{B})$ exactly when there exists a positive integer $r$ such that  there is a surjective homomorphism from the power structure $\mathcal{A}^r$ to $\mathcal{B}$. 

We note that this condition is already necessary to guarantee containment of $\Pi_2$-$\mathrm{CSP}(\mathcal{A}) \subseteq \Pi_2$-$\mathrm{CSP}(\mathcal{B})$. Thus we can say on finite structures that positive Horn collapses to its $\Pi_2$ fragment. If the sizes of the structures $\mathcal{A}$ and $\mathcal{B}$ are $|A|$ and $|B|$, respectively, then we may take $r:=|A|^{|B|}$. Thus to decide whether $\mathrm{QCSP}(\mathcal{A}) \subseteq \mathrm{QCSP}(\mathcal{B})$, it suffices to verify whether or not there is a surjective homomorphism from $\mathcal{A}^{|A|^{|B|}}$ to $\mathcal{B}$. This provides a decision procedure for the model containment problem with non-deterministic double-exponential time complexity.

Keisler had already established in \cite{Keisler65} that a necessary and sufficient condition for countable $\mathcal{A}$ and $\mathcal{B}$ to satisfy  $\mathrm{QCSP}(\mathcal{A}) \subseteq \mathrm{QCSP}(\mathcal{B})$ is a surjective homomorphism from $\mathcal{A}^\omega$ to $\mathcal{B}$. Thus our result can be seen as complementing his with a bound on $r$ in the case that $\mathcal{A}$ and $\mathcal{B}$ are finite. Keisler's result holds also for infinite structures, and appears as part of a much more general result (with remarkably elegant proof) whose principal object of study is in fact the Horn fragment of first-order logic. His methods are typical of those used in (Classical) Model Theory: a back-and-forth argument making use of the benevolent properties of infinity. In the transfinite case his results rely on the Continuum Hypothesis.  

It is possible to prove our model containment result using the traditional back-and-forth proof method. However, we show that the positive Horn collapse to $\Pi_2$ is not observable via the back-and-forth because it does not hold on suitably chosen infinite structures. Indeed, if one allows for $\mathcal{B}$ to be not quite finite, but still $\omega$-categorical (while $\mathcal{A}$ remains finite) then already the $\Pi_2$ collapse fails.

We demonstrate that our combinatorial result extends to give the $\Pi_2$ collapse in the case where $\mathcal{B}$ remains finite but $\mathcal{A}$ is $\omega$-categorical, but, as mentioned, show that it can not be extended to the case where $\mathcal{A}$ is finite and $\mathcal{B}$ is $\omega$-categorical. 

We demonstrate a near-matching lower bound to the exponent of $\mathcal{A}$, by giving a sequence of structures $\mathcal{A}$ together with an fixed $\mathcal{B}$, $|B|=2$, such that there is a surjective homomorphism from $\mathcal{A}^r$ to $\mathcal{B}$ only when $r \geq |A|$. This is only a square away from the upper bound $|A|^{|B|}$ $=|A|^2$. The simplest structures we use have a growing signature, but we detail a fixed finite signature variant with the same properties.

The Classical Decision Problem, known also as Hilbert's \emph{Entscheidungsproblem}, is the question, given a first-order sentence $\varphi$, whether $\varphi$ is true in all models (is logically valid) or, dually, is true in some model (is satisfiable). It is well-known that this problem is undecidable in general. The entailment problem for first-order logic asks, given sentences $\varphi$ and $\psi$, whether we have the logical validity of $\varphi \rightarrow \psi$ (denoted $\models \varphi \rightarrow \psi$). The equivalence problem is defined similarly, with $\rightarrow$ substituted by $\leftrightarrow$. Both problems are easily seen to be equivalent to the Classical Decision Problem, and are therefore undecidable. A great literature exists on decidable and undecidable cases of the Classical Decision Problem, particularly under restrictions of quantifier prefixes and (arity and number of) relation and function symbols -- see the monograph \cite{CDP}. However, for certain natural fragments of first-order logic, it seems the entailment and equivalence problems are not well-studied. The query containment problem is closely related to the entailment problem, but with truth in all finite models substituted for truth in all models. Query containment problems are fundamental to many aspects of database systems, including query optimisation, determining independence of queries and rewriting queries using views. The query containment problem for first-order logic is also undecidable.

The sentence containment problem for the CSP -- \mbox{a.k.a.} the query containment problem for primitive positive logic -- is the question, given primitive positive sentences $\varphi$ and $\psi$, whether, for all finite structures $\mathcal{A}$, $\mathcal{A} \models \varphi$ implies $\mathcal{A} \models \psi$ (\mbox{i.e.} $\models_\mathrm{fin} \varphi \rightarrow \psi$). It is easily seen that this problem is decidable and NP-complete, in fact it is an instance of the homomorphism problem (equivalently, the CSP itself). It is also easy to demonstrate, in this case, that the condition of finiteness is irrelevant. That is, $\models_\mathrm{fin} \varphi \rightarrow \psi$ if, and only if, $\models \varphi \rightarrow \psi$. Thus we have here the decidability and NP-completeness of the entailment problem for primitive positive logic.

The second part of this paper is motivated by the sentence containment problem for the QCSP -- \mbox{a.k.a.} the query containment problem for positive Horn -- that is, given positive Horn sentences $\varphi$ and $\psi$, to determine whether $\models_\mathrm{fin} \varphi \rightarrow \psi$. In this case it is not clear as to whether this coincides with the condition of entailment, $\models \varphi \rightarrow \psi$.
Our principle contribution here is to give a decision procedure, with triple-exponential time complexity, for the entailment problem, \mbox{i.e.} the problem to determine, for two pH-sentences $\varphi$ and $\psi$, whether $\models \varphi \rightarrow \psi$. Since primitive positive sentences are positive Horn, it follows from the comments of the previous paragraph that this entailment problem is NP-hard.

We will make particular use of a certain canonical model for the sentence $\varphi$, built on the Herbrand universe of terms derived from Skolem functions over a countably infinite set of (new) constants. Herbrand models are commonplace in algorithmic results on logical validity and equivalence in both first-order logic (e.g. \cite{Kozen81}) and logic programming (e.g. \cite{lifschitz01strongly,EiterFTW07,EiterFW07}). However, our method differs significantly from those in the citations.

We also prove that the related entailment problem for positive logic -- even without equality -- is undecidable. Since the difference between positive Horn and positive logic is simply the addition of disjunction, we suggest that our decidability result is somehow tight.

In the last part of the paper, we go on to consider canonical representatives of classes of the equivalence relation $\sim_{\mathrm{pH}}$ induced by $\mathcal{A} \sim_{\mathrm{pH}} \mathcal{B}$ iff  $\mathrm{QCSP}(\mathcal{A}) = \mathrm{QCSP}(\mathcal{B})$. The similar relation for pp-logic always has a unique minimal element, the so-called \emph{core}, which is minimal with regard to both cardinality and induced submodel. The consideration of only cores simplifies considerably many CSP classifications, and is tantamount to considering the related polymorphism algebra to be idempotent. The situation for QCSP we show to be somewhat murkier, and we contrast positive Horn in this regard to primitive positive logic, positive equality-free logic and positive logic. We introduce the \emph{Q-cores} and show that, although their behaviour is difficult to pin down, this notion is able to greatly simplify known QCSP classification. We comment finally on the role of idempotency in the algebraic method applied to QCSPs.

This paper is organised as follows. After the preliminaries, we address the QCSP model containment problem in Section~\ref{sec:RHS}. Then we address the positive Horn entailment problem in Section~\ref{sec:LHS}. Finally, we expose the nature of Q-cores in Section~\ref{sec:Q-cores}. We conclude with some final remarks and open questions.

\textbf{Related work}. This paper is an expanded journal version of \cite{LICS2008} together with the most significant parts of \cite{QCores}. In particular, the discussion of the $\omega$-categorical case is new to this paper.

For a structure $\mathcal{A}$, let $\langle \mathcal{A} \rangle_{\mathrm{pH}}$ be the set of relations positive Horn definable on $\mathcal{A}$. Let $\langle \mathcal{A} \rangle_{\Pi_2\mbox{-}\mathrm{pH}}$ be that subset of these relations that are already definable in the $\Pi_2$ fragment. It follows from \cite{OxfordAndHubie} (see \cite{HubieSIGACT} for details though not a proof) that, for all $\mathcal{A}$,  $\langle \mathcal{A} \rangle_{\mathrm{pH}}$ and  $\langle \mathcal{A} \rangle_{\Pi_2\mbox{-}\mathrm{pH}}$ actually coincide. This phenomenon is related to our $\Pi_2$ collapse.

\section{Global Preliminaries}

Let $\sigma$ be a fixed, finite relational signature. If $\mathcal{B}$ is a $\sigma$-structure, then its domain is denoted $B$ and the cardinality of that domain $|B|$. The stipulation that $\sigma$ contains no constants is purely for technical convenience, as we will wish to consider structures over the expanded signature $\sigma \cup C_m$, where $C_m$ is a set of (an arbitrary number) $m$ constant symbols. These constants will be used specifically to name elements of the structure that correspond to the evaluation of universal variables. Structures over the expanded $\sigma \cup C_m$ will be written in Fraktur, $\mathfrak{B}$, whereupon their $\sigma$-reducts become $\mathcal{B}$, in the natural way. For $R \in \sigma$ and a $\sigma$-structure $\mathcal{B}$, we sometimes write $R(\overline{x}) \in \mathcal{B}$ to indicate $\mathcal{B} \models R(\overline{x})$.

A \emph{homomorphism} from $\mathcal{A}$ to $\mathcal{B}$ is a function $h:A \rightarrow B$ that preserves positive relations. That is, for $R$ a $p$-ary relation symbol of $\sigma$, if $(x_1,\ldots,x_p) \in R^\mathcal{A}$ then $(h(x_1),\ldots,h(x_r)) \in R^\mathcal{B}$. A homomorphism $h:\mathfrak{A} \rightarrow \mathfrak{B}$ must also preserve the constants, i.e. if $x=c_i^\mathfrak{A}$ then $h(x)=c_i^\mathfrak{B}$.
Existence of a homomorphism (resp., surjective homomorphism) from $\mathcal{A}$ to $\mathcal{B}$ is denoted $\mathcal{A} \rightarrow \mathcal{B}$ (resp., $\mathcal{A} \surhom \mathcal{B}$). If both $\mathcal{A} \rightarrow \mathcal{B}$ and $\mathcal{B} \rightarrow \mathcal{A}$, then we describe $\mathcal{A}$ and $\mathcal{B}$ as \emph{homomorphically equivalent}. If $f:A\rightarrow B$ is a function, and $A'\subseteq A$ then we denote by $\mathrm{Im}(f)[A']$ the image of $A'$ under $f$ (i.e. $\{ f(x) : x \in A' \}$). When $A'$ is omitted, it is considered to be the whole set $A$.

A first-order (fo) sentence $\varphi$ is \emph{positive} if it contains no instances of negation and is \emph{positive Horn} (pH) if, further, it contains no instances of disjunction. Thus, pH involves precisely $\forall$, $\exists$ and $\wedge$ (and $=$, a topic we will return to later in the paper). If we further forbid universal quantifiers then we arrive at a sentence that is \emph{primitive positive} (pp). A priori, pp and pH sentences may contain equalities, though it is easy to see these may be propagated out in all but trivial cases (a topic we will return to later). It is clear that a positive (resp., pH, pp) sentence may be put in the prenex normal form
\[\varphi \ := \ \forall \tuple{x}_1 \exists \tuple{y}_1 \ldots \forall \tuple{x}_k \exists \tuple{y}_k \ P(\tuple{x}_1,\tuple{y}_1,\ldots,\tuple{x}_k,\tuple{y}_k), \]
where $P$ is positive (resp., a conjunction of atoms). If $\varphi$ contains only variables $\tuple{x}_1$ and $\tuple{y}_1$ (i.e. one quantifier alternation) then it is said to be $\Pi_2$; if $\varphi$ contains only (the existential) variables $\tuple{x}_1$ then it is said to be $\Sigma_1$.
The \emph{quantified constraint satisfaction problem} $\mathrm{QCSP}(\mathcal{A})$ has
\begin{itemize}
\item Input: a positive Horn sentence $\varphi$.
\item Question: does $\mathcal{A} \models \varphi$?
\end{itemize}
If $\varphi$ is restricted to being $\Pi_2$ (resp., $\Sigma_1$) then the resulting problem is $\Pi_2$-$\mathrm{CSP}(\mathcal{A})$ (resp., $\mathrm{CSP}(\mathcal{A})$). We identify a problem with the set of its yes-instances in the obvious way. The \emph{model containment problem} for QCSP takes as input two finite structures $\mathcal{A}$ and $\mathcal{B}$, and has as its yes-instances those pairs for which $\mathrm{QCSP}(\mathcal{A})$ $\subseteq \mathrm{QCSP}(\mathcal{B})$. The model containment problem for CSP and $\Pi_2$-CSP is defined analogously.


Let $\varphi$ be a prenex sentence of the form $\forall \tuple{x}_1 \exists \tuple{y}_1 \ldots \forall \tuple{x}_k \exists \tuple{y}_k \ P(\tuple{x}_1,\tuple{y}_1,\ldots,\tuple{x}_k,\tuple{y}_k)$, and let $\mathcal{A}$ be a finite structure. We will identify a variable tuple $\tuple{x}$ with its underlying set of variables. The \emph{$\varphi$-game on $\mathcal{A}$} is a two-player game that pitches
\emph{Universal} (male) against \emph{Existential} (female). 
The game goes as follows.
For $1 \leq i \leq k$ ascending: 
\begin{itemize}
\item for every variable in $\tuple{x}_i$, Universal chooses an element in
  $A$: i.e. he gives a function $f_{\forall_i}:\tuple{x}_i\rightarrow A$;
  and, 
\item for every variable in $\tuple{y}_i$, Existential chooses an element in
  $A$: i.e. she gives a function $f_{\exists_i}:\tuple{y}_i \rightarrow A$.
\end{itemize}
Existential wins if, and only if, 
\[ \mathcal{A} \models P(f_{\forall_1}(\tuple{x}_1),f_{\exists_1}(\tuple{y}_1),\ldots,f_{\forall_k}(\tuple{x}_k),f_{\exists_k}(\tuple{y}_k)), \]
where $f(\tuple{x})$ is the natural pointwise action of $f$ on the coordinates of $\tuple{x}$. 

A \emph{strategy} $\varepsilon:=(\varepsilon_1, \ldots,\varepsilon_k)$ for Existential (resp., $\upsilon:=(\upsilon_1,\ldots,\upsilon_{k})$ for Universal) tells
her (resp., him) how to play a variable tuple given what has been played
before. That is, $\varepsilon_l$ is a function from 
$A^{(\tuple{x}_1 \cup \tuple{y}_1 \cup \ldots \cup \tuple{x}_{l-1})} \times \tuple{y}_l$
to $A$ and $\upsilon_l$ is a function from 
$A^{(\tuple{x}_1 \cup \tuple{y}_1 \cup \ldots \cup \tuple{x}_{l-1} \cup \tuple{y}_{l-1})} \times \tuple{x}_l$
to $A$ (note that elements of $A^{(\tuple{x}_1 \cup \tuple{y}_1 \cup \ldots \cup \tuple{x}_{l-1})}$ and $A^{(\tuple{x}_1 \cup \tuple{y}_1 \cup \ldots \cup \tuple{x}_{l-1} \cup \tuple{y}_{l-1})}$ are themselves functions specifying how the game was played on the previous variable tuples). 
A strategy for Existential is \emph{winning} if it beats all possible strategies of Universal. The $\varphi$-game on $\mathcal{A}$ is nothing other than a model-checking (Hintikka) game, and it is straightforward to verify that Existential has a winning strategy if, and only if, $\mathcal{A} \models \varphi$. In the case where $P$ is a conjunction of atoms, then the winning condition may be recast as their being a homomorphism to $\mathcal{A}$ from the structure specified by the atomic conjunction $P$ (this construction will be resurrected in the sequel). 

Given two $\sigma \cup C_m$-structures $\mathfrak{A}$ and $\mathfrak{B}$, we define their \emph{direct} (or categorical) \emph{product} $\mathfrak{A} \bigotimes \mathfrak{B}$ to have domain $A \times B$ and relations
$((x_1,y_1),\ldots,(x_{a_i},y_{a_i})) \in R_i^{\mathfrak{A} \bigotimes \mathfrak{B}}$ iff $(x_1,\ldots,x_{a_i}) \in R^\mathfrak{A}$ and $(y_1,\ldots,y_{a_i}) \in R_i^\mathfrak{B}$. The constant $c_i^{\mathfrak{A} \bigotimes \mathfrak{B}}$ is the element $(x_i,y_i)$ s.t. $x_i=c_i^ \mathfrak{A}$ and $y_i=c_i^\mathfrak{B}$. Note that the operator $\bigotimes$ is associative and commutative, up to isomorphism. Bearing this in mind, $\mathfrak{A}^m$ indicates the power structure $\mathfrak{A} \bigotimes \cdots \bigotimes \mathfrak{A}$, from $m$ copies of $\mathfrak{A}$, where $m$ may be any cardinal.

The \emph{orbit} of an $n$-tuple $(a_1,\ldots,a_n)$ of elements in a structure $\mathcal{A}$ is the set 
\[ \{ (a'_1,\ldots,a'_n) : \mbox{ there is an automorphism of $\mathcal{A}$ mapping $(a_1,\ldots,a_n)$ to $(a'_1,\ldots,a'_n)$.} \}\]
A countably infinite structure is said to be \emph{$\omega$-categorical} if it is the unique countable model of its first-order theory. It is known by the theorem of Engeler, Ryll-Nardzewsky and Svenonius (see \cite{Hodges}) that a structure that is $\omega$-categorical has a finite number of orbits of $n$-tuples, for each $n$. This is one of several ways in which an $\omega$-categorical structure may be said to be ``finite'' in its behaviour.

The \emph{Continuum Hypothesis} (CH) is the assertion that there is no cardinal strictly between $\omega$ and $2^\omega$, i.e. $\omega+=2^\omega$.

\section{The QCSP Model Containment Problem}
\label{sec:RHS}

The following lemma is a restriction of the well-known fact that surjective homomorphisms preserve positive formulae (see, e.g., \cite{Hodges}) -- we sketch the proof for the sake of completeness.
\begin{lem}\label{RHS:lem:pos-pres}
For all $\mathcal{A}$ and $\mathcal{B}$, if $\mathcal{A} \surhom \mathcal{B}$ then $\mathrm{QCSP}(\mathcal{A})  \subseteq \mathrm{QCSP}(\mathcal{B})$.
\end{lem}
\proof[(Sketch)]
If $s:\mathcal{A} \rightarrow \mathcal{B}$ is a surjective homomorphism, then let $s^{-1}:B \rightarrow A$ be s.t. $s \circ s^{-1}$ is the identity on $B$. Let $\varphi$ be a pH sentence of the form $\forall \tuple{x}_1 \exists \tuple{y}_1 \ldots \forall \tuple{x}_k \exists \tuple{y}_k $ $P(\tuple{x}_1,\tuple{y}_1,\ldots,\tuple{x}_k,\tuple{y}_k)$. Given a winning strategy $\varepsilon$ for Existential in the $\varphi$-game on $\mathcal{A}$, we build a winning strategy $\varepsilon'$ for her in the $\varphi$-game on $\mathcal{B}$. For $1 \leq i \leq k$, let $g$ be a mapping from $(\tuple{x}_1 \cup \tuple{y}_1 \cup \ldots \cup \tuple{x}_{i-1})$ to $B$ and let $y$ be a variable of $\tuple{y}_i$. We set $\varepsilon'_i(g,y):= s \circ \varepsilon_i( s^{-1} \circ g, y )$. The result follows from the positivity of $P$
\qed

\begin{figure*}
  \centering
  \resizebox{!}{2cm}{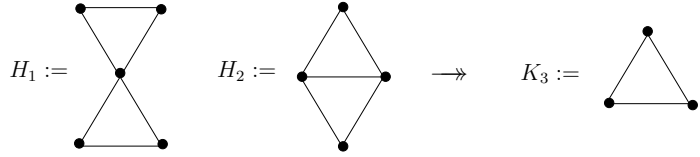}  
  \caption{Two graphs and a homomorphic image.}
  \label{fig:H1H2K3}
\end{figure*}

\begin{figure*}
  \centering
$
\xymatrix{
0  \ar@{<->}[dr]  \ar@{<->}[ddr]  \ar@{<->}[drr] \ar@/_5pc/@{<->}[ddrr] &
1 \ar@{<->}[dr] \ar@{<->}[dl] \ar@{<->}[ddr] \ar@{<->}[ddl] & 
2 \ar@{<->}[dl] \ar@{<->}[dll] \ar@{<->}[ddl] \ar@{<->}[ddl] \\
0 \ar@{<->}[urr] \ar@{<->}[drr] \ar@{<->}[dr] &
1 \ar@{<->}[dr] \ar@{<->}[dl] & 
2 \ar@{<->}[dl] \\
3 \ar@{<->}[urr]  \ar@/_5pc/@{<->}[uurr] &
1 & 
2  \\
}
$ \hspace{1cm} $\xymatrix{\\ \surhom\\ \\}$ \hspace{1cm}
$
\xymatrix{
& 0  \ar@{<->}[dr]  \ar@{<->}[dl] & \\
1  \ar@{<->}[rr] & & 2 \\
& 3  \ar@{<->}[ur]  \ar@{<->}[ul] & \\
}
$
  \caption{Surjective homomorphism from ${\mathcal{K}_3}^2$ to $\mathcal{H}_2$.}
  \label{fig:k23toh2}
\end{figure*}
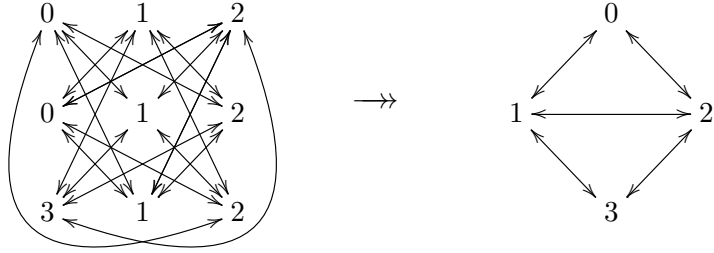

\begin{ex}
\label{ex:graphs}
Consider the graphs drawn in Figure~\ref{fig:H1H2K3}. Both $\mathcal{H}_1$ and $\mathcal{H}_2$ have a surjective homomorphism to $\mathcal{K}_3$; therefore we can derive both $\mathrm{QCSP}(\mathcal{H}_1)  \subseteq \mathrm{QCSP}(\mathcal{K}_3)$ and $\mathrm{QCSP}(\mathcal{H}_2)  \subseteq \mathrm{QCSP}(\mathcal{K}_3)$.
\end{ex}
\begin{lem}
\label{lem:product:strategy}
For all $\mathcal{A}$ and $r>0$, $\mathrm{QCSP}(\mathcal{A}) \subseteq \mathrm{QCSP}(\mathcal{A}^r)$.
\end{lem}
\proof
Let $\varphi$ be a pH sentence of the form $\forall \tuple{x}_1 \exists \tuple{y}_1 \ldots \forall \tuple{x}_k \exists \tuple{y}_k$ $P(\tuple{x}_1,\tuple{y}_1,\ldots,$ $\tuple{x}_k,\tuple{y}_k)$. Let $\varepsilon$ be a winning strategy for Existential in the $\varphi$-game on $\mathcal{A}$. The product strategy $\varepsilon^r$ for Existential in the $\varphi$-game on $\mathcal{A}^r$ is defined as follows.
For $1 \leq i \leq k$, let $g$ be a mapping from $(\tuple{x}_1 \cup \tuple{y}_1 \cup \ldots \cup \tuple{x}_{i-1})$ to $A^r$ and let $y$ be a variable of $\tuple{y}_i$. We set $\varepsilon^r_i(g,y):=$
$ (\varepsilon_i(\pi_1\circ g,y),\ldots,\varepsilon_i(\pi_r\circ g,y))$, 
where $\pi_1,\ldots,\pi_r$ denote the natural projections from $A^r$ to $\mathcal{A}$. That $\varepsilon^r$ is a winning strategy for Existential in the $\varphi$-game on $\mathcal{A}^r$ follows from the fact that $P$ is a conjunction of atoms (because every atom must have been true in every one of the $r$ components).
\qed
\begin{remarks}
Lemma~\ref{lem:product:strategy} holds with the same proof for any ordinal exponent $r$.
While Lemma~\ref{RHS:lem:pos-pres} holds for all positive sentences (not just pH), Lemma~\ref{lem:product:strategy} does not hold for positive sentences in general. Consider the directed $1$-path $\mathcal{DP}_1$, \mbox{i.e.} the digraph with vertex set $\{1,2\}$ and edge set $\{(1,2)\}$. Take $\varphi:=\forall x\exists y E(x,y) \vee E(y,x)$. $\mathcal{DP}_1 \models \varphi$ but ${\mathcal{DP}_1}^2 \notmodels \varphi$. In Figure~\ref{fig:H1H2K3}, ${\mathcal{K}_3}^2 \surhom \mathcal{H}_2$ (homomorphism exhibited in Figure~\ref{fig:k23toh2}), so we can deduce that $\mathcal{K}_3$ and $\mathcal{H}_2$ agree on all pH sentences.
\end{remarks}

\noindent We note the following which essentially appears in \cite{Keisler65}.
\begin{thm}[\cite{Keisler65}]
Let $\mathcal{B}$ be finite and $\mathcal{A}$ of any cardinality. Then $\mathrm{QCSP}(\mathcal{A})  \subseteq \mathrm{QCSP}(\mathcal{B})$ iff $\mathcal{A}^\omega \surhom \mathcal{B}$.
\end{thm}

\subsection{Combinatorial characterisation}
\label{sec:comb}
\begin{thm}\label{theo:main:result}
  Let $\mathcal{A}$ and $\mathcal{B}$ be finite $\sigma$-structures. The following are equivalent.
  \begin{enumerate}[label=\Roman*.]
  \item[I.] $\mathcal{A}^{|A|^{|B|}} \surhom \mathcal{B}$.
  \item[II.] There exists $r<\omega$ s.t. $\mathcal{A}^r \surhom \mathcal{B}$.
  \item[III.] $\mathrm{QCSP}(\mathcal{A})  \subseteq \mathrm{QCSP}(\mathcal{B})$.
  \item[IV.] $\Pi_2\mbox{-}\mathrm{CSP}(\mathcal{A})  \subseteq \Pi_2\mbox{-}\mathrm{CSP}(\mathcal{B})$.
  \end{enumerate}
\end{thm}
\noindent We now set out to prove Theorem~\ref{theo:main:result}, essentially through combinatorial means. 
Recall the signature $\sigma \cup C_m$, where $C_m:=\{c_1,\ldots,c_m\}$. We will associate $C_m$ with $[m]:=\{1,\ldots,m\}$, in the natural way. Given a mapping $\lambda$ from $[m]$ to a structure $\mathcal{A}$, we
write $\mathfrak{A}_\lambda$ for the $\sigma \cup C_m$-structure induced naturally by $\mathcal{A}$ and the
interpretation of the constant symbols given by $\lambda$. Let $A^{[m]}$
denote the set of  all possible interpretations. We call \emph{Superprodukt} the $\sigma \cup C_m$-structure $\mathfrak{A}^{|A|^m}:=\bigotimes_{\lambda \in
  A^{[m]}} \mathfrak{A}_\lambda$. Note that this is well-defined since $\otimes$ is associative and commutative, up to isomorphism; and does not produce a clash of notation, as no structure $\mathfrak{A}$ has been defined. From its definition it is clear to see that $\mathfrak{A}^{|A|^m}$ is some kind of enriched power structure of $\mathcal{A}$ (indeed, it shares a  domain with $\mathcal{A}^{|A|^m}$).

There is a natural correspondence between $\Pi_2$ pH sentences $\varphi$ with $m$ universally quantified variables and
$\sigma \cup C_m$-structures. Recall $\varphi$ is of the form $\forall \tuple{x}_1 \exists \tuple{y}_1 \ P(\tuple{x}_1,\tuple{y}_1)$, where $\tuple{x}_1:=(x^1_1,\ldots,x^m_1)$ and $P$ is a conjunction of atoms.
From $\varphi$, we build the $\sigma \cup C_m$-structure $\mathfrak{D}_\varphi$ in the following way. The elements of $\mathfrak{D}_\varphi$ are the variables of $\varphi$, and the relation tuples of $\mathfrak{D}_\varphi$ are exactly the facts of the conjunction $P(\tuple{x}_1,\tuple{y}_1)$ (indeed if all the quantifiers of $\varphi$ were switched to being existential then one would obtain the so-called canonical query -- see \cite{KolaitisVardiBook05} -- of the structure $\mathcal{D}_\varphi$). Finally, the elements $x^1_1,\ldots,x^m_1$ interpret the constants $c_1,\ldots,c_m$. 
Conversely, given a $\sigma \cup C_m$-structure
$\mathfrak{D}$, we build the $\Pi_2$ pH sentence $\varphi_\mathfrak{D}$ as follows. The variables of $\varphi_\mathfrak{D}$ are the elements of $\mathfrak{D}$, and the quantifier-free part of $\varphi_\mathfrak{D}$ is the conjunction of the facts of $\mathfrak{D}$. Finally, the variables (whose elements interpreted the constants) $c_1,\ldots,c_m$ are universally quantified, while all other variables are existentially quantified (to the inside of the universal quantification).  This correspondence is essentially bijective, and is illustrated in the following example. 
\begin{ex}
\label{ex:one}
$\varphi := \forall x^1_1, x^2_1, x^3_1 \ \exists y^1_1,y^2_1,y^3_1,y^4_1 \ E(y^1_1,x^1_1) \wedge
E(x^1_1,y^2_1) \wedge E(x^1_1,y^3_1) \wedge E(y^3_1,y^2_1) \wedge E(y^4_1,x^2_1)
\wedge E(x^3_1,y^4_1)$. 

The sentence $\varphi$, depicted on the left, gives rise to the $\sigma \cup C_3$-structure $\mathfrak{D}_\varphi$, depicted on the right. The existential variables and their corresponding elements are not labelled. 

\vspace{0.5cm}
\hspace{2cm}
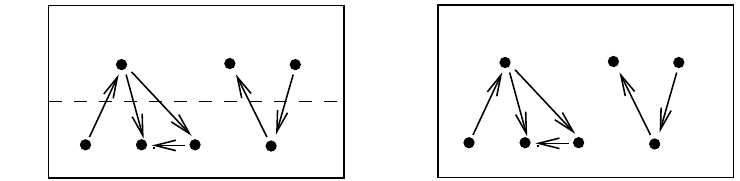
\end{ex}
\begin{lem}
\label{RHS:thm:methodology}
  Let $\varphi$ be of the form $\forall \tuple{x}_1 \exists \tuple{y}_1 \ P(\tuple{x}_1,\tuple{y}_1)$, where $P$ is a conjunction of positive atoms and $\tuple{x}_1:=(x^1_1,\ldots,x^m_1)$. Let $\mathfrak{D}_\varphi$ be $\varphi$'s corresponding
  $\sigma \cup C_m$-structure. The following are equivalent:
  \begin{enumerate}[label=(\roman*)]
  \item $\mathcal{A} \models \varphi$
  \item $\mathfrak{D}_\varphi \homm \mathfrak{A}^{|A|^m}$
  \end{enumerate}
\end{lem}
\proof
  $\mathcal{A} \models \varphi$ iff for every mapping $f_{\forall_1}$ from $\tuple{x}_1$ to
  $A$, there exists a mapping $f_{\exists_1}$ from $\tuple{y}_1$ to $A$ such that
$\mathcal{A} \models P(f_{\forall_1}(\tuple{x}_1), f_{\exists_1}(\tuple{y}_1))$.
From the definition, this is equivalent to there existing a homomorphism from $\mathfrak{D}_\varphi$ to $\mathfrak{A}_\lambda$, for every $\lambda \in A^{[m]}$ (indeed, when $\lambda$ coincides with $f_{\forall_1}$, under the natural substitution of the domain $[m]$ by $(x^1_1,\ldots,x^m_1)$, then  $f_{\forall_1} \cup f_{\exists_1}$ provides the homomorphism).
 By construction of $\mathfrak{A}^{|A|^m}$ as a product of such
  $\mathfrak{A}_\lambda$, we have equivalently that there exists a homomorphism
  from $\mathfrak{D}_\varphi$ to $\mathfrak{A}^{|A|^m}$.
\qed
\proof[(of Theorem~\ref{theo:main:result})]
I $\Rightarrow$ II  is trivial. II $\Rightarrow$ III follows from Lemmas~\ref{RHS:lem:pos-pres} and \ref{lem:product:strategy}.  III $\Rightarrow$ IV is trivial. 

It remains to prove  IV $\Rightarrow$ I. Assume $\Pi_2\mbox{-}\mathrm{CSP}(\mathcal{A})  \subseteq \Pi_2\mbox{-}\mathrm{CSP}(\mathcal{B})$. Consider $\varphi_{\mathfrak{A}^{|A|^{|B|}}}$. Clearly, $\mathcal{A} \models \varphi_{\mathfrak{A}^{|A|^{|B|}}}$, by the upward direction of Lemma~\ref{RHS:thm:methodology}. It follows from our assumption that $\mathcal{B} \models \varphi_{\mathfrak{A}^{|A|^{|B|}}}$. Let $f_{\forall_1}:\tuple{x}_1 \rightarrow B$ be any surjective function and $f_{\exists_1}:\tuple{y}_1 \rightarrow B$ be given according to a winning strategy for Existential in the $\varphi_{\mathfrak{A}^{|A|^{|B|}}}$-game on $\mathcal{B}$. But now $f_{\forall_1} \cup f_{\exists_1}$ gives a surjective homomorphism from $\mathcal{A}^{|A|^{|B|}}$ to $\mathcal{B}$ which proves our result.
\qed
\begin{ex}\label{ex:bip}
  Consider an undirected bipartite graph with at least one edge
  $\mathcal{G}$ and $\mathcal{K}_2$ the graph that consists of a single double-edge.
  There is a surjective homomorphism from $\mathcal{G}$ to $\mathcal{K}_2$.
  Note also that $\mathcal{K}_2\otimes \mathcal{K}_2 = \mathcal{K}_2 \uplus\mathcal{K}_2$ (where $\uplus$ stands for disjoint union)
  which we write as $2\cdot \mathcal{K}_2$. Thus, ${\mathcal{K}_2}^j=2^{j-1}\cdot \mathcal{K}_2$ (as $\otimes$
  distributes over $\uplus$). Hence, if $\mathcal{G}$ has no isolated element and
  $m$ edges there is a surjective homomorphism from ${\mathcal{K}_2}^{1+\lceil \log_2
    m \rceil}$ to $\mathcal{G}$. It follows from Theorem~\ref{theo:main:result}) that  $\mathrm{QCSP}(\mathcal{K}_2)  = \mathrm{QCSP}(\mathcal{G})$.
\end{ex}

\subsection{Complexity}
\label{RHS:Complexity}

Having established a combinatorial characterisation for the QCSP model containment problem, we make the following observation as to its complexity -- as can be seen the twin bounds are far from tight.
\begin{thm}
The model containment problem for QCSP, that is the problem which, given finite structures $\mathcal{A}$ and $\mathcal{B}$, decides whether $\mathrm{QCSP}(\mathcal{A}) \subseteq \mathrm{QCSP}(\mathcal{B})$ is 1.) in nondeterministic double-exponential time, and 2.) NP-hard (under polynomial-time reductions).
\end{thm}
\proof
Membership of nondeterministic double-exponential time follows from Theorem~\ref{theo:main:result} by building $\mathcal{A}^{|A|^{|B|}}$ and guessing a surjective homomorphism to $\mathcal{B}$ (which can easily be verified as such in double-exponential time). NP-hardness follows by a reduction from the problem \emph{graph $3$-colourability}, as we will demonstrate.

Let $\mathcal{K}_1$ and $\mathcal{K}_3$ be the (irreflexive) $1$- and $3$-clique, respectively. That is, $\mathcal{K}_1$ is a single loopless vertex and $\mathcal{K}_3$ is the triangle. 
Recall $3.\mathcal{K}_1$ is the graph $\mathcal{K}_1 \uplus \mathcal{K}_1 \uplus \mathcal{K}_1$. It is well-known that $\mathcal{G}$ is $3$-colourable iff $\mathcal{G} \homm \mathcal{K}_3$. It is easy to see that this is equivalent to $(\mathcal{G} \uplus 3\cdot \mathcal{K}_1) \surhom \mathcal{K}_3$. We claim that this is equivalent to the existence of an $r$ s.t. $(\mathcal{G} \uplus 3\cdot \mathcal{K}_1)^r \surhom \mathcal{K}_3$. 
To see this, use first the fact that $\mathcal{G}$ is an induced substructure of $(\mathcal{G} \uplus 3.\mathcal{K}_1)^r$ (note that for any $r$, $\mathcal{D}$ is a substructure of $\mathcal{D}^r$) to derive the existence of a homomorphism from $\mathcal{G}$ to $\mathcal{K}_3$. This homomorphism can be used in turn to construct a surjective homomorphism from $\mathcal{G} \uplus 3.\mathcal{K}_1$ to $\mathcal{K}_3$.
The result now follows from Theorem~\ref{theo:main:result}. 
\qed

\subsection{Extending Theorem~\ref{theo:main:result}}
\label{sec:omega-cat}

The exponent $|A|^{|B|}$ of Theorem~\ref{theo:main:result} corresponds to the number of functions $\lambda:[|B|]\rightarrow \mathcal{A}$. Suppose $\lambda_0$ and $\lambda_1$ are distinct functions s.t. there is an automorphism of $\mathcal{A}$ mapping $(\lambda_0(1),\ldots,\lambda_0(|B|))$ to $(\lambda_1(1),\ldots,\lambda_1(|B|))$, then it can be seen that we do not in fact need both of these in the Superprodukt $\bigotimes_{\lambda \in
  A^{[m]}} \mathfrak{A}_\lambda$, as the $\lambda_0$th and
$\lambda_1$th components are isomorphic. A first upper bound on the
exponent is therefore the number of distinct orbits of $|B|$-tuples in
$\mathcal{A}$, and we will now see how this will enable us to derive a
version of Theorem~\ref{theo:main:result} when $\mathcal{B}$ is finite
and $\mathcal{A}$ is potentially infinite. The application of
K\"onig's Lemma in the following proof is based on that in
\cite{BodirskyNesetrilJLC}.
\newpage
\begin{thm}\label{theo:omega-cat}
  Let $\mathcal{A}$ be $\omega$-categorical and $\mathcal{B}$ a finite $\sigma$-structure. The following are equivalent.
  \begin{enumerate}[label=\Roman*.]
  \item[I.] $\mathcal{A}^{\omega} \surhom \mathcal{B}$.
  \item[II.] There exists $r<\omega$ s.t. $\mathcal{A}^r \surhom \mathcal{B}$.
  \item[III.] $\mathrm{QCSP}(\mathcal{A})  \subseteq \mathrm{QCSP}(\mathcal{B})$.
  \item[IV.] $\Pi_2\mbox{-}\mathrm{CSP}(\mathcal{A})  \subseteq \Pi_2\mbox{-}\mathrm{CSP}(\mathcal{B})$.
  \end{enumerate}
\end{thm}
\proof
Again: I $\Rightarrow$ II  is trivial. II $\Rightarrow$ III follows from Lemmas~\ref{RHS:lem:pos-pres} and \ref{lem:product:strategy}. III $\Rightarrow$ IV is trivial. 

In Theorem~\ref{theo:main:result} we proved IV $\Rightarrow$ I, but here we prefer IV $\Rightarrow$ II (knowing that II $\Rightarrow$ I is trivial).  Assume IV.  Let the number of distinct orbits of $|B|$-tuples in $\mathcal{A}$ be $z$.
Enumerate the countable domain $A^z$ by $\alpha_1,\alpha_2,\ldots$. For $m \geq 1$, consider the set $\Gamma_m$ of finite partial homomorphisms from $\mathcal{A}^z$ restricted to $\{\alpha_1,\ldots,\alpha_m\}$ to $\mathcal{B}$. That such always exist is attested by the fact that the canonical query of$\mathcal{A}^z$ restricted to $\{\alpha_1,\ldots,\alpha_m\}$, itself a pp-sentence and true on  $\mathcal{A}^z$, is also by assumption true on $\mathcal{B}$. We introduce an equivalence relation on $\Gamma_m$ whereby $g\sim h$ if there is an automorphism $\mathit{aut}$ of $\mathcal{A}^z$ s.t. $g=f \circ \mathit{aut}$. These equivalence classes will form nodes of a forest in which there are edges joining (the equivalence class of) a finite partial homomorphism $f$ on domain $\{\alpha_1,\ldots,\alpha_m\}$ with (the equivalence class of) its extension on domain $\{\alpha_1,\ldots,\alpha_{m+1}\}$. By assumption, this forest has nodes representing all finite domains that ultimately cover $\mathcal{A}^z$. It has a  finite number of trees, since there is a bounded number of $|B|$-types in $\mathcal{A}^z$ (which is $\omega$-categorical since $\mathcal{A}$ is), and each tree is infinite. Further, each tree is finitely branching, since the number of distinct orbits of $n$-tuples in $\mathcal{A}^z$ is finite. It follows from K\"onig's Lemma that there is an infinite branch in each tree that gives a homomorphism from $\mathcal{A}^z$ to $\mathcal{B}$.
\qed

\subsection{Limit of the method}
\label{sec:limit}

We will now show that we do not observe the $\Pi_2$ collapse, that manifests in Theorem~\ref{theo:main:result}, in the general case. A fuller statement of the result of \cite{Keisler65} would be as follows.
\begin{thm}[\cite{Keisler65}]
\label{thm:keisler2}
Assume the CH. Let $\mathcal{B}$ be of cardinality $\omega+$ and saturated (or finite), and let $\mathcal{A}$ be of cardinality at most $\omega+$. Then $\mathrm{QCSP}(\mathcal{A})  \subseteq \mathrm{QCSP}(\mathcal{B})$ iff $\mathcal{A}^\omega \surhom \mathcal{B}$.
\end{thm}
\noindent We will establish the following.
\begin{prop}
\label{prop:counterexample}
There is a finite $\mathcal{A}$ and $\omega$-categorical $\mathcal{B}$ s.t. $\Pi_2\mbox{-}\mathrm{CSP}(\mathcal{A})  \subseteq \Pi_2\mbox{-}\mathrm{CSP}(\mathcal{B})$ but not $\mathrm{QCSP}(\mathcal{A})  \subseteq \mathrm{QCSP}(\mathcal{B})$.
\end{prop}
\noindent Assuming the infinite part of Theorem~\ref{thm:keisler2} is not vacuous -- i.e. assuming the CH -- the $\mathcal{B}$ can be substituted by a saturated elementary extension of cardinality $2^\omega$ (see \cite{Marker}). So, assuming the CH, Theorem~\ref{thm:keisler2} is actually untrue with pH substituted by $\Pi_2$-pH. We will begin by establishing the following.
\begin{lem}
\label{lem:NandQ}
$\Pi_2\mbox{-}\mathrm{CSP}(\mathbb{N};\leq)  \subseteq \Pi_2\mbox{-}\mathrm{CSP}(\mathbb{Q};\leq)$.
\end{lem}
\proof
For $\varphi$ positive Horn, and given a winning strategy $\varepsilon$ for Existential in the $\varphi$-game on $(\mathbb{N};\leq)$, we will build a winning strategy $\varepsilon'$ for her in the $\varphi$-game on $(\mathbb{Q};\leq)$. Let $g:\tuple{x}_1\rightarrow \mathbb{Q}$ be given and let $m$ be the least common multiple of the denominators in $\mathrm{Im}(g)$.
Set $\varepsilon'_1(g,y):=1/m(\varepsilon_1(mg,my))$ (where $mg$ indicates the function of multiplication by $m$ concatenated on $g$).
\qed
\noindent Let $\mathcal{DP}^*_1$ be the digraph with vertex set $\{1,2\}$ and edge set $\{(1,1),(1,2),(2,2)\}$.
\begin{lem}
\label{lem:lately}
There is a surjective homomorphism $s$ from ${\mathcal{DP}^*_1}^\omega$ to $([m];\leq)$.
\end{lem}
\proof
Indeed, we give a surjective homomorphism from ${\mathcal{DP}^*_1}^{m-1}$ to $([m];\leq)$. Set $s(1,\ldots,1)$ $=1$ and $s(2,\ldots,2)=m$. Now, for $(x_1,\ldots,x_{k-1})$ of the form $x_1,\ldots,x_i=2$ and $x_{i+1}=1$, set $s(x_1,\ldots,x_{k-1})=i+1$.
\qed
\noindent Note that one can even argue there is a surjective homomorphism from ${\mathcal{DP}^*_1}^{\log m}$ to $([m];\leq)$.
\begin{lem}
\label{lem:DP1andN}
$\Pi_2\mbox{-}\mathrm{CSP}({\mathcal{DP}^*_1}^\omega)  \subseteq \Pi_2\mbox{-}\mathrm{CSP}(\mathbb{N};\leq)$.
\end{lem}
\proof
For $\varphi$ positive Horn, and given a winning strategy $\varepsilon$ for Existential in the $\varphi$-game on ${\mathcal{DP}^*_1}^\omega$, we will build a winning strategy $\varepsilon'$ for her in the $\varphi$-game on $(\mathbb{N};\leq)$. Let $g:\tuple{x}_1\rightarrow \mathbb{N}$ be given and let $m$ be the maximum of $\mathrm{Im}(g)$. Let $s$ a surjective homomorphism $s$ from ${\mathcal{DP}^*_1}^\omega$ to $([m];\leq)$ as given by Lemma~\ref{lem:lately}.
Set $\varepsilon'_1(g,y):=s(\varepsilon_1((s^{-1}\circ g,y)))$, where inverse images under $s$ are chosen arbitrarily.
\qed
\proof[Proof of Proposition~\ref{prop:counterexample}]
That $\Pi_2\mbox{-}\mathrm{CSP}(\mathcal{DP}^*_1;\leq)  \subseteq \Pi_2\mbox{-}\mathrm{CSP}(\mathbb{Q};\leq)$ follows from Lemmas~\ref{RHS:lem:pos-pres}, \ref{lem:product:strategy}, \ref{lem:NandQ} and \ref{lem:DP1andN}. However, the positive Horn sentence $\exists x \forall y \ x\leq y$ holds on the former, but not on the latter (when the edge relation of $\mathcal{DP}^*_1$ is identified with an order).
\qed
\noindent Proposition~\ref{prop:counterexample} may also be seen as limiting the methods used in the previous section.

\subsection{Lower bounds on the exponent}
\label{sec:lowerbound}

The Example~\ref{ex:bip} of bipartite graphs gives us a lower bound on the exponent that we now seek to improve.
Let $\sigma:=\langle U_1,\ldots,U_k\rangle$ be a signature involving $k$ unary relations. Let $\mathcal{A}_k$ be the $\sigma$-structure with domain $A_k:=\{1,\ldots,k\}$ where $U_i:=A_k \setminus \{i\}$, for each $i \in [k]$. Let $\mathcal{B}$ be the $\sigma$-structure  with domain $B:=\{0,1\}$, where $1 \in U_i$ and $0 \notin U_i$, for all $i \in [k]$. It is clear that ${\mathcal{A}_k}^{k} \surhom \mathcal{B}$ (rainbow elements of the for $(x_1,\ldots,x_k)$ where $|\{x_1,\ldots,x_k\}|=k$ can map to $0$) while ${\mathcal{A}_k}^{k-1} \nosurhom \mathcal{B}$. For $\mathcal{A}:=\mathcal{A}_k$, this gives us a lower bound on the exponent of $|A|$ where the upper bound would give $|A|^{|B|}=|A|^2$.

It is not too demanding to construct a finite signature variant of this. Consider the signature $\sigma:=\langle E,U\rangle$ involving a binary relation $E$ and a unary relation $U$. Let $\mathcal{A}_k$ be the directed cycle on $k$ vertices, such that all except one of these vertices is in the relation $U$. Let $\mathcal{B}$ have domain $B:=\{0,1\}$ with $E:=\{(0,0),(1,1)\}$ and $U:=\{1\}$. This has the property that ${\mathcal{A}_k}^{k} \surhom \mathcal{B}$ while ${\mathcal{A}_k}^{k-1} \nosurhom \mathcal{B}$. For $\mathcal{A}:=\mathcal{A}_k$, this once again gives us a lower bound on the exponent of $|A|$ where the upper bound would give $|A|^{|B|}=|A|^2$.

\section{The Entailment Problem}
\label{sec:LHS}

For a simpler exposition, we will assume throughout this section that all pH sentences have strict quantifier alternation, i.e. are of the form
\[\varphi \ := \ \forall x_1 \exists y_1 \ldots \forall x_k \exists y_k \ P(x_1,y_1,\ldots,x_k,y_k), \]
where $P$ is a conjunction of positive atoms. Of course, any pH sentence may be readily converted to an equivalent sentence in this form by the introduction of dummy variables. If $P$ contains any atomic instance $x_i=x_j$ ($i \neq j$) or $y_i=x_j$ ($i < j$) then we describe $\varphi$ as \emph{degenerate}. It is clear that all models of a degenerate $\varphi$ are of cardinality $1$, and that there is a finite set of normalised $\sigma$-structures over the domain $\{1\}$. It follows that, if $\varphi$ is degenerate, we may establish directly whether $\models \varphi \rightarrow \psi$ by evaluating $\psi$ over all normalised models of $\varphi$. 

Note that instances of equality in a non-degenerate $\varphi$ may be propagated out by substitution. In order to answer the question $\models \varphi \rightarrow \psi$ in general, we will wish to build a canonical model of $\varphi$. Henceforth, we will assume that $\varphi$ (but not necessarily $\psi$) contains no instances of equality.

\subsubsection*{The Canonical Model}

Let $\varphi$ be a pH sentence of the form 
\[\forall x_1 \exists y_1 \ldots \forall x_k \exists y_k\ P(x_1,y_1,\ldots,x_k,y_k)\,.\]
We consider $k$ to be the \emph{depth} of $\varphi$, denoted $\depth(\varphi)$. We wish to build a \emph{canonical model} of $\varphi$, and we shall do this via its Skolem normal form. Let $F:=\{f_1,\ldots,f_k\}$ be a set of function symbols, in which the arity of $f_i$ is $i$. Let 
\[ \skolem(\varphi) :=\forall x_1 \ldots \forall x_k \ P(x_1,f_1(x_1),\ldots,x_k,f_k(x_1,\ldots,x_k))\]
be the derivative sentence over the signature $\sigma \cup F$. Each atom of $P$ induces what we designate a \emph{quantified atom} in $\skolem(\varphi)$. It is well-known that the models of $\varphi$ and $\skolem(\varphi)$ are intimately related, indeed they are identical up to the additional interpretation of the new function symbols of $F$.

If $\alpha$ is a positive integer, let $C_\alpha:=\{c_1,\ldots,c_\alpha\}$; if $\alpha:=\omega$, let $C_\alpha:=\{c_1,\ldots \}$. 
Define $T_\varphi(C_\alpha)$ to be the set of (closed) \emph{terms} obtained from all compositions of the functions of $F$ on themselves and on the constants of $C_\alpha$. The \emph{rank} of a term $t \in T_\varphi(C_\alpha)$, denoted $\rank(t)$, is the maximum nesting depth of its function symbols; $C_\alpha$ is precisely that subset of $T_\varphi(C_\alpha)$ of terms of rank $0$. Define $T^m_\varphi(C_\alpha)$ to be the subset of $T_\varphi(C_\alpha)$ induced by terms whose rank is $\leq m$.
Note that $T_\varphi(C_\alpha)$ is exactly the domain of the \emph{term algebra} of $\sigma \cup F \cup C_\alpha$ (see, e.g., \cite{Hodges}).

Considering all instantiations of $x_1,\ldots,x_k$ by the terms of $T_\varphi(C_\alpha)$, we see that $\skolem(\varphi)$ becomes an infinite set of positive atoms $\Phi$, exactly the instantiations of the quantified atoms of $\skolem(\varphi)$. These immediately give rise to a canonical (sometimes known as Herbrand) model of $\skolem(\varphi)$ over the domain  $T_\varphi(C_\alpha)$ in the standard way (see, e.g., \cite{Hodges}); we denote this model $\mathcal{T}_\varphi(C_\alpha)$ (\mbox{i.e.} with calligraphic $T$). Note that $\Phi$ is the positive (Robinson) diagram of $\mathcal{T}_\varphi(C_\alpha)$. Rather sloppily, we will consider $\mathcal{T}_\varphi(C_\alpha)$ to be at once a $\sigma$-structure (a bona fide model of $\varphi$) and a $\sigma \cup F \cup C_\alpha$-structure -- this should cause no confusion. By further abuse of nomenclature, we will also continue referring to the elements of $T_\varphi(C_\alpha)$ as `terms' and elements of $C_\alpha \subseteq T_\varphi(C_\alpha)$ as `constants'. Let $\mathcal{T}^m_\varphi(C_\alpha)$ be the \emph{truncation} (submodel) of $\mathcal{T}_\varphi(C_\alpha)$ induced by the domain $T^m_\varphi(C_\alpha)$. Note that $\mathcal{T}^m_\varphi(C_\alpha)$ is generally not a model of $\varphi$; however, the following is immediate from the construction.
\begin{fact}
For all $\alpha$, $\mathcal{T}_\varphi(C_\alpha) \models \varphi$.
\end{fact}
\begin{ex}
\label{ex:main}
Let $\sigma:=\langle E \rangle$ contain a single binary relation (i.e. $\sigma$-structures are digraphs).
Let $\varphi:=\forall x \forall z \exists y \ E(x,y) \wedge E(y,z)$. In this case,\footnote{The reader may notice that $\varphi$ is not in the correct form as it fails to have strict alternation of quantifiers. While the introduction of a dummy existential quantifier (and consequent dummy unary Skolem function in $\skolem(\varphi)$) would rectify this, it would also make the example rather hard to follow.} 
\[ \skolem(\varphi):= \forall x \forall z \ E(x,f(x,z)) \wedge E(f(x,z),z).\]
The quantified atoms of $\skolem(\varphi)$ are 
\[
\begin{array}{c}
\forall x \forall z E(x,f(x,z)) \mbox{ and} \\
\forall x \forall z E(f(x,z),z). 
\end{array}
\]

\noindent The following are depictions of the truncations $\mathcal{T}^2_\varphi(C_1)$ and $\mathcal{T}^1_\varphi(C_2)$, respectively.

\[
\xymatrix{
& c \ar@/_/[ddl] \ar@/_/[d] & \\
& f(c,c) \ar@/_/[u] \ar@/_/[d] \ar[dr] & \\
f(c,f(c,c)) \ar[ur] & f(f(c,c),f(c,c)) \ar@/_/[u] & f(f(c,c),c) \ar@/_/[uul] \\
}
\]

\[
\xymatrix{
c_1 \ar@/_/[d] \ar[dr] & & & c_2 \ar@/^/[d] \ar[dl] \\
f(c_1,c_1) \ar@/_/[u] & f(c_1,c_2) \ar@/^/[urr] & f(c_2,c_1) \ar@/_/[ull] & f(c_2,c_2) \ar@/^/[u] \\
}
\]

\end{ex}

\subsubsection{A Surjective Diagram Lemma}

Let $\varphi$ be a pH sentence, $F$ its associated set of Skolem functions and $\skolem(\varphi)$ its Skolem normal form. The canonical model $\mathcal{T}_\varphi(C_\omega)$, with a countably infinite set of constants, plays a key role in our discourse.
The following is a variant of the Diagram Lemma (see, e.g., \cite{Hodges}).
\begin{lem}
\label{lemma:surjective-diagram}
Let $\varphi$ be a pH sentence. Then, for all countable (not necessarily infinite) structures $\mathcal{B}$, if $\mathcal{B} \models \varphi$ then there is a surjective homomorphism $h:\mathcal{T}_\varphi(C_\omega) \rightarrow \mathcal{B}$ s.t. $h(C_\omega)=B$.
\end{lem}
\proof
Let $b_1,\ldots$ be an enumeration of the elements of $\mathcal{B}$. Let $\mathfrak{B}$ be the expansion of $\mathcal{B}$, over the signature $\sigma \cup C_\omega$ s.t. the elements $b_1,\ldots$ interpret the constants $c_1,\ldots$ (if $\mathcal{B}$ is finite interpret all remaining constants as, e.g., $b_1$). Since $\varphi$ contains no constants, $\mathfrak{B} \models \varphi$. It follows that there is a further expansion $\overline{\mathfrak{B}}$ over the signature $\sigma \cup F \cup C_\omega$, s.t. $\overline{\mathfrak{B}} \models  \skolem(\varphi)$

Considering $\mathcal{T}_\varphi(C_\omega)$ as a $\sigma \cup F \cup C_\omega$-structure, we now uncover the canonical function $h:\mathcal{T}_\varphi(C_\omega) \rightarrow \overline{\mathfrak{B}}$. Each $t \in T_\varphi(C_\omega)$ is a syntactic term over $F \cup C_\omega$. Set $h(t)$ to be the element (which interprets) $t$ in $\overline{\mathfrak{B}}$.

The function $h$ is manifestly a homomorphism, since $\overline{\mathfrak{B}} \models \skolem(\varphi)$ (actually, it is also unique).  

By once again considering $\mathcal{T}_\varphi(C_\omega)$ to be a $\sigma$-structure, we see that $h$ is a surjective homomorphism from $\mathcal{T}_\varphi(C_\omega)$ to $\mathcal{B}$, s.t. $h(C_\omega)=B$.
\qed

\subsection{Characterisation}
\label{LHS:sec:characterisation}

We are now in a position to derive a model-theoretic characterisation for $\models \varphi \rightarrow \psi$.
\begin{thm}
\label{thm:methodology}
Let $\varphi$ and $\psi$ be pH sentences. The following are equivalent:
\begin{itemize}
\item[$\bullet$] $\models \varphi \rightarrow \psi$, i.e. $\varphi \rightarrow \psi$ is logically valid, and
\item[$\bullet$] $\mathcal{T}_\varphi(C_\omega) \models \psi$.
\end{itemize}
\end{thm}
\proof
(Downwards.) Since $\models \varphi \rightarrow \psi$, we derive $\mathcal{T}_\varphi(C_\omega) \models \varphi \rightarrow \psi$, whence, since $\mathcal{T}_\varphi(C_\omega) \models \varphi$, we derive $\mathcal{T}_\varphi(C_\omega) \models \psi$.

(Upwards.) Suppose $\mathcal{T}_\varphi(C_\omega) \models \psi$ and, for some $\mathcal{A}$, we have $\mathcal{A} \models \varphi$. If $\mathcal{A}$ is infinite and uncountable, then we apply the  Downward L\"{o}wenheim-Skolem Theorem to find another, countable, model $\mathcal{A}'$ that agrees with $\mathcal{A}$ on all first-order sentences. It follows from Lemma~\ref{lemma:surjective-diagram} that there is a surjective homomorphism $h:\mathcal{T}_\varphi(C_\omega) \rightarrow \mathcal{A}'$. It now follows from  Lemma~\ref{RHS:lem:pos-pres} that $\mathcal{A}' \models \psi$ and hence so does $\mathcal{A}$.
%
%
\qed

\subsubsection{Restricting Universal's Play}

Now let $\varphi$ be a pH sentence of which $\mathcal{T}_\varphi(C_\alpha)$ is a canonical model. Let $\psi$ be a pH sentence of the form $\forall x_1 \exists y_1 \ldots \forall x_l \exists y_l$ $Q(x_1,y_1,$ $\ldots,x_l,y_l)$. The \emph{$\psi$-rel-game on $\mathcal{T}_\varphi(C_\alpha)$} is defined similarly to the $\psi$-game on $\mathcal{T}_\varphi(C_\alpha)$, except Universal is now restricted to playing elements of $C_\alpha \subseteq T_\varphi(C_\alpha)$. In this case, Existential has a winning strategy in the $\psi$-rel-game on $\mathcal{T}_\varphi(C_\alpha)$ iff $\mathcal{T}_\varphi(C_\alpha) \models$ 
\[ \forall x_1 \in C_\alpha \exists y_1 \ldots \forall x_l \in C_\alpha \exists y_l \ Q(x_1,y_1,\ldots,x_l,y_l),\]
that is, if $\mathcal{T}_\varphi(C_\alpha)$ models $\psi$ with the universal variables relativised to $C_\alpha$.
\begin{prop}
\label{prop:restrict-universal}
Let $\varphi$ and $\psi$ be pH sentences, with $\mathcal{T}_\varphi(C_\alpha)$ a canonical model of $\varphi$. Then,
Existential has a winning strategy in the $\psi$-game on $\mathcal{T}_\varphi(C_\alpha)$, i.e. $\mathcal{T}_\varphi(C_\alpha) \models \psi$, iff Existential has a winning strategy in the $\psi$-rel-game on $\mathcal{T}_\varphi(C_\alpha)$.
\end{prop}
\proof
The forward direction is trivial. The backward direction may be proved in a similar manner to Lemma~\ref{RHS:lem:pos-pres}, given that Lemma~\ref{lemma:surjective-diagram} provides us with a surjective endomorphism $s:\mathcal{T}_\varphi(C_\omega) \rightarrow \mathcal{T}_\varphi(C_\omega)$ s.t. $s(C_\omega)=T_\varphi(C_\omega)$.
\qed

\subsubsection{Substitution Lemmas}

Given a term $t \in T_\varphi(C_\omega)$ one may consider the various \emph{subterms} of which it is composed. For example, the term $f(f(c_1,c_2),f(f(c_1,c_1),$ $c_2))$ of rank $3$ contains both $c_2$ and $f(c_1,c_1)$ as subterms. We will talk of a term $t$ as \emph{containing} the constants that are its subterms.
We adopt the notation $t[t'/t'']$ to denote the term obtained by replacing, in $t$, all instances of $t'$ by $t''$ (nota bene $t'$ by $t''$).

Consider terms $t_1,t_2,\ldots,t_r,t',t'' \in T_\varphi(C_\omega)$.
Suppose that $R(t_1,t_2,\ldots,t_r)$ holds in the canonical model $\mathcal{T}_\varphi(C_\omega)$; might it always be the case that $R(t_1[t'/t''],$ $t_2[t'/t''],$ $\ldots,t_r[t'/t''])$ holds in $\mathcal{T}_\varphi(C_\omega)$? The answer is no; for example, in the case of digraphs, if $E(c,f(c)) \in \mathcal{T}_\varphi(C_\omega)$, then we have no reason to conclude that $E(c,c) \in \mathcal{T}_\varphi(C_\omega)$, even though the latter corresponds to $E(c[f(c)/c],f(c)[f(c)/c])$. However, we can make substitutions subject to certain rules, as the following lemmata attest. 

\begin{lem}[Substitution of terms of distinct rank]
\label{lem:sub-terms}
Let $R$ be a $p$-ary relation symbol of $\sigma$, and consider $t_1,\ldots,t_p,t' \in T_\varphi(C_\omega)$ s.t. $\rank(t')$ is distinct from each of $\rank(t_1)$, \ldots, $\rank(t_p)$. For all terms $t''$, if $R(t_1,\ldots,t_p) \in \mathcal{T}_\varphi(C_\omega)$ then $R(t_1[t'/t''],\ldots,t_p[t'/t'']) \in \mathcal{T}_\varphi(C_\omega)$.
\end{lem}
\proof
Consider the quantified atom of $\skolem(\varphi)$ that caused $R(t_1,\ldots,t_p)$ to be in $\mathcal{T}_\varphi(C_\omega)$ (via its instantiation in the positive diagram $\Phi$). It must have been of the form
\[ \forall \overline{z}_1 \ldots \forall \overline{z}_p \ R(g_1(\overline{z}_1),\ldots,g_p(\overline{z}_p)), \]
where $\overline{z}_1, \ldots, \overline{z}_p$ are not required to be disjoint, and each $g_i$ is either
\begin{itemize}
\item the identity $\iota$ (in which case $\overline{z}_i$ is a singleton) or 
\item some $f_j \in F$ (in which case $\overline{z}_i$ is a $j$-tuple).
\end{itemize}
Since $t'$ is distinct in rank from each of $t_1,\ldots,t_p$, it can be easily seen that all occurrences of $t'$ in the $t_1,\ldots,t_p$ of  $R(t_1,\ldots,t_p)$ must have come from occurrences of $t'$ in the instantiations of the variables $\overline{z}_1, \ldots, \overline{z}_p$. It follows that the related instantiation $\overline{z}_1[t'/t''], \ldots, \overline{z}_p[t'/t'']$ yields $R(t_1[t'/t''],\ldots,t_p[t'/t''])$, and the result follows.
\qed

\begin{lem}[Substitution of constants]
\label{lem:sub-consts}
Let $R$ be a $p$-ary relation symbol of $\sigma$, consider $t_1,\ldots,t_p \in T_\varphi(C_\omega)$ and $c,c' \in C_\omega$. If $R(t_1,\ldots,t_p) \in \mathcal{T}_\varphi(C_\omega)$ then $R(t_1[c/c'],\ldots,t_p[c/c']) \in \mathcal{T}_\varphi(C_\omega)$.
\end{lem}
\proof
Similar to the previous lemma.
\qed
Let $\pi:C_\omega \rightarrow C_\omega$ be some (partial) bijection. For a term $t \in T_\varphi(C_\omega)$, let $\pi(t)$ be the term obtained by simultaneously switching each constant $c_i$ for $\pi(c_i)$, in the obvious manner.
\begin{lem}[Permutation of constants]
\label{lem:aut}
Let $R$ be a $p$-ary relation symbol of $\sigma$, and consider $t_1,\ldots,t_p \in T_\varphi(C_\omega)$.
Then, $R(t_1,\ldots,t_p) \in \mathcal{T}_\varphi(C_\omega)$ iff $R(\pi(t_1),\ldots,$ $\pi(t_p)) \in$ $\mathcal{T}_\varphi(C_\omega)$.
\end{lem}
\proof
It is evident from the construction that, for each permutation $\pi$, $\mathcal{T}_\varphi(C_\omega)$ has an automorphism that maps each term $t$ to $\pi(t)$.
\qed

The structure $\mathcal{T}_\varphi(C_\alpha)$ has the useful property that any finite substructure $\mathcal{A} \subseteq \mathcal{T}_\varphi(C_\alpha)$ has a homomorphism to the truncation $\mathcal{T}^{|A|}_\varphi(C_\alpha)$. In fact, we are able to derive a stronger property. Call a partial function $f:{T}_\varphi(C_\alpha) \rightarrow {T}_\varphi(C_\alpha)$ \emph{constant-conservative} if, for all $t \in {T}_\varphi(C_\alpha)$, $f(t)$ contains no constants that are not contained in $t$.
\begin{lem}
\label{lem:AhomT}
For $\mathcal{A} \subseteq \mathcal{T}_\varphi(C_\alpha)$, there is a constant-conservative homomorphism $\mathcal{A} \homm \mathcal{T}^{|A|}_\varphi(C_\alpha)$.
\end{lem}
\noindent The general idea of the proof is, in the (worst) case that the terms of $\mathcal{A}$ have distinct ranks, that they can still all be mapped to the first ${|A|}$ ranks in a way that preserves the rank-order. The proof uses Lemma~\ref{lem:sub-terms} in order to explain what we do when a rank has been `missed out' in $\mathcal{A}$. Indeed, when a rank has been missed out, then we may reduce the rank of all higher terms in the rank-order, in an almost arbitrary way, while preserving homomorphism. However, to ensure that the homomorphism is constant-conservative, we reduce rank in a more particular manner. 
\proof
Let $t_1,\ldots,t_{|A|}$ be the elements of $\mathcal{A}$ ordered by non-decreasing rank. If the maximal rank is $> {|A|}$ then there exists some $t_i \in A$ of rank $r$ s.t. no $t \in A$ is of rank $r-1$, and $t_i$ is of the form $f_j(s_1,\ldots,s_j)$ for some terms $s_1,\ldots,s_j$ of which (at least) one is of rank $r-1$. Suppose one that is of rank $r-1$ is $s_m$. Pick any subterm $s'_m$ of $s_m$ of rank $r-2$. Let $\mathcal{A}'$ be that substructure of $\mathcal{T}_\varphi(C_\alpha)$ derived by substituting $s'_m$ for $s_m$ in all the terms of $A$. Clearly this substitution is constant-conservative. We claim that the function from $\mathcal{A}$ to $\mathcal{A}'$ induced by this substitution is a homomorphism, whereupon we may iterate the above reasoning until the obtained structure has maximal rank $\leq {|A|}$.

(Proof that $\mathcal{A} \homm \mathcal{A}'$.) 
Consider the elements $t_1,\ldots,t_{|A|}$ of $\mathcal{A}$ and the natural map that takes them to $t_1[s_m/s'_m], \ldots,t_{|A|}[s_m/s'_m]$ in $\mathcal{A}'$. We will demonstrate that this is a homomorphism. Let $R$ be a $p$-ary relation symbol of $\sigma$. 
Suppose $R(t_{\lambda_1},\ldots,t_{\lambda_p}) \in \mathcal{A} \subseteq \mathcal{T}_\varphi(C_\alpha) \subseteq \mathcal{T}_\varphi(C_\omega)$, by Lemma~\ref{lem:sub-terms} we have $R(t_{\lambda_1}[s_m/s'_m],\ldots,$ $t_{\lambda_p}[s_m/s'_m]) \in \mathcal{T}_\varphi(C_\omega)$, whereupon the result follows (since $\mathcal{A}'$ is an induced substructure of $\mathcal{T}_\varphi(C_\alpha) \subseteq \mathcal{T}_\varphi(C_\omega)$).

It remains to argue that the iterative procedure we have given terminates, \mbox{i.e.} that eventually we end up in a situation in which the elements of $\mathcal{A}$ have ranks $\leq |A|$. We show why the iteration of our procedure ultimately produces a model in which all terms are of rank $\leq |A|$.
Let $\mu(\mathcal{A}):=\sum_{t \in A} \rank(t)$, and suppose $A$ contains a term $t$ of rank $>|A|$.
We claim that after $\leq k$ iterations of our procedure (where $k=\depth(\varphi)$) we must obtain a $\mathcal{A}'$ s.t. $\mu(\mathcal{A}') < \mu(\mathcal{A})$, whereupon convergence of our procedure is implied.
Suppose, as before, that $\mathcal{A}$ contains no term of rank $r-1$, but contains some $t_i$, of rank $r$ and of the form $f_j(s_1,\ldots,s_j)$, s.t. $t_i$ contains $z \leq j \leq k$ subterms of rank $r-1$ (i.e. $s_{i_1},\ldots,s_{i_z}$ are of rank $r-1$, and $s_m \in \{s_{i_1},\ldots,s_{i_z}\}$). Either $\mathcal{A}$ is s.t. $\mu(\mathcal{A}') < \mu(\mathcal{A})$ or $\mathcal{A}'$ also contains no term of rank $r-1$, but contains $t_i[s_l/s'_l]$, of rank $r$ s.t. $t_i[s_l/s'_l]$ contains $z' < z$ subterms of rank $r-1$. The result follows.
\qed

\subsubsection{Restricting Existential's Play}

Proposition~\ref{prop:restrict-universal} tells us that we may consider Universal's play restricted to the set $C_\alpha$ in the $\psi$-game on $\mathcal{T}_\varphi(C_\alpha)$. Now we detail how we may make a certain assumption about Existential's play, without affecting her ability to win.

Let $\varphi,\psi$ be pH sentences, with $\psi$ of the form $\forall x_1 \exists y_1 \ldots \forall x_l \exists y_l \ Q(x_1,y_1,\ldots,x_l,y_l)$, and let $\mathcal{T}_\varphi(C_\alpha)$ be a canonical model of $\varphi$. Define the \emph{$\psi$-rel-cc-game on $\mathcal{T}_\varphi(C_\alpha)$} as the $\psi$-rel-game on $\mathcal{T}_\varphi(C_\alpha)$ but now restrict Existential to only playing terms $t$ containing constants that Universal has already played (the cc abbreviates constant-conservative). In other words, if Universal has played $c_{j_1},\ldots,c_{j_i}$ for variables $x_1,\ldots,x_i$, then Existential must play some $t \in T_\varphi(\{c_{j_1},\ldots,c_{j_i}\})$ for $y_i$. Legitimate strategies for Existential in this game will be termed \emph{constant-conservative}. Winning strategies for Existential in the $\psi$-rel-cc-game on $\mathcal{T}_\varphi(C_\alpha)$ are central to our discourse.

Consider the $\psi$-rel-game (resp., $\psi$-rel-cc-game) on the truncation $\mathcal{T}^{m}_\varphi(C_\alpha) \subseteq \mathcal{T}_\varphi(C_\alpha)$ defined in the obvious way.
\begin{prop}
\label{prop:cc}
Let $\varphi,\psi$ be pH sentences, with $\psi$ of the form 
\[\forall x_1 \exists y_1 \ldots\ \forall x_l \exists y_l \ Q(x_1,y_1,\ldots,x_l,y_l)\,.\]
 The following are equivalent.
  \begin{enumerate}[label=(\roman*)]
\item  Existential has a winning strategy in the $\psi$-rel-game on $\mathcal{T}_\varphi(C_\omega)$.
\item  Existential has a winning strategy in the $\psi$-rel-cc-game on $\mathcal{T}_\varphi(C_\omega)$.
\item Existential has a winning strategy in the $\psi$-rel-cc-game on $\mathcal{T}_\varphi(C_l)$.
\item Existential has a winning strategy in the $\psi$-rel-cc-game on $\mathcal{T}^{l^{l+2}}_\varphi(C_l)$.
\end{enumerate}
\end{prop}
\proof We break the proof into a cyclic system of implications.
\paragraph{$(i \Rightarrow ii)$}
Consider a \emph{game tree} $\mathscr{G}_\varepsilon$ for the $\psi$-rel-game on $\mathcal{T}_\varphi(C_\omega)$ under Existential strategy $\varepsilon$. $\mathscr{G}_\varepsilon$ is an out-tree, branching on all possible Universal moves over $C_\omega$ when Existential plays according to $\varepsilon$. The branching factor of $\mathscr{G}_\varepsilon$ from the root to the leaves is alternately $\omega$ and $1$, and the distance from the root to the leaves is $2l$. The nodes at distance $2i-1$ (resp., $2i$) from the root are labelled with Universal's (resp., Existential's) $i$th move. The root is unlabelled. If $\varepsilon$ is a winning strategy, then when we read off valuations for $x_1,y_1,\ldots,x_l,y_l$ on a path, we will always have $\mathcal{T}_\varphi(C_\omega) \models Q(x_1,y_1,\ldots,x_l,y_l)$.

We will modify $\mathscr{G}_\varepsilon$ inductively from the root to the leaves, in such a way as to ultimately enforce that Existential's moves are constant-conservative while keeping her strategy winning. The property $(*)$ that we will maintain is that, at distance $\leq 2i$ from the root, there is no node $\lambda$ labelled by an Existential play $t$ containing a constant $c$ that Universal has not played on the path from the root to $\lambda$. When $i=0$ this is clearly true; and when $i=2l$ we have that Existential's play was always constant-conservative. 

Suppose the inductive hypothesis $(*)$ holds at distance $\leq 2i$ from the root. While there is a node $\lambda$, at distance $2(i+1)$ from the root, labelled by an Existential play $t$ containing a constant $c$ that Universal has not played on the path from the root to $\lambda$, we undertake the following procedure.
\begin{itemize}
\item Remove all subtrees beyond $\lambda$ whose roots are labelled with Universal plays $c$. 
\item Pick a constant $c'$ that has been already played by Universal on the path from the root to $\lambda$, and substitute all terms $t$ labelling nodes in the subtree rooted at $\lambda$ with $t[c/c']$.
\end{itemize}
It follows from Lemma~\ref{lem:sub-consts} that this modified game tree still represents a winning strategy for Existential, \emph{so long as Universal never plays $c$ beyond node $\lambda$}. 

Now consider all missing subtrees corresponding to Universal plays of $c$ after $\lambda$. These follow Existential plays at nodes $\lambda_1:=\lambda$, $\lambda_2$, \ldots, $\lambda_{(l-i-1)}$ at distances $0,2,\ldots,2(l-i-1)$ beyond $\lambda$. For each $r \in \{0,1,\ldots,2(l-i-1)\}$, consider what Universal plays for $x_{i+1+r}$:
\begin{itemize}
\item Pick some next Universal play that is a constant $c''$ s.t. $c''$ has not appeared on the path from the root to $\lambda_r$ (such a constant must exist since only a finite number of constants can be mentioned on any path). 
\item Take the bijection $\pi:C_\omega \rightarrow C_\omega$ that swaps $c$ and $c''$. Duplicate the subtree corresponding to the choice $c''$ (i.e. rooted at the node labelled $c''$ immediately after $\lambda$) but reset all the node labels $t$ to $\pi(t)$. Now reintroduce this subtree as the choice $c$ (immediately after $\lambda$). 
\end{itemize}
Since neither $c''$ nor $c$ is mentioned before $\lambda_r$, it follows from Lemma~\ref{lem:aut} that this modified game tree still represents a winning strategy for Existential.

An example for case $(i \Rightarrow ii)$ follows the remainder of the proof.

\paragraph{$(ii \Rightarrow iii)$}
Existential may use the same winning strategy in the $\psi$-rel-cc-game on $\mathcal{T}_\varphi(C_l)$ as she used in the $\psi$-rel-cc-game on $\mathcal{T}_\varphi(C_\omega)$. This is because her play is constant-conservative.

\paragraph{$(iii \Rightarrow iv)$}
Consider a winning strategy $\varepsilon$ in the $\psi$-rel-cc-game on $\mathcal{T}_\varphi(C_l)$. We will construct a winning strategy $\varepsilon'$ for her in the $\psi$-rel-cc-game on $\mathcal{T}^{l^{l+2}}_\varphi(C_l)$.
Recall $x_1,\ldots,x_l$ are the ordered universal variables of $\psi$; there are at most $l^l$ ways in which they may be, in order, played on to the set $C_l$. This means that Existential needs at most $l\cdot l^l$ elements of $\mathcal{T}_\varphi(C_l)$ to beat any strategy of Universal. This means that there is a substructure $\mathcal{A} \subseteq \mathcal{T}_\varphi(C_l)$ that contains at most $l\cdot l^l$ elements other than those of $C_l$ s.t. Existential has the winning strategy $\varepsilon$ in the $\psi$-rel-cc-game on $\mathcal{A}$. Note that $|A| \leq l+l\cdot l^l \leq l^{l+2}$.

Let $h:\mathcal{A} \homm \mathcal{T}^{l^{l+2}}_\varphi(C_l)$ be a (constant-conservative) homomorphism, as guaranteed by Lemma~\ref{lem:AhomT}. It follows that $\varepsilon':=h \circ \varepsilon$ suffices.

\paragraph{$(iv \Rightarrow i)$}
Suppose Existential has a winning strategy $\varepsilon$ in the $\psi$-rel-cc-game on $\mathcal{T}^{l^{l+2}}_\varphi(C_l)$, we will construct a (constant-conservative) winning strategy $\varepsilon'$ for her in the $\psi$-rel-game on $\mathcal{T}_\varphi(C_\omega)$. At the $j$th round, Existential has in mind a partial bijection $\pi_j:C_\omega \rightarrow C_\omega$.\enlargethispage{2\baselineskip}

Universal plays first, with some constant $c_{i_1}$ for $x_1$. Existential sets $\pi_1:=c_1\mapsto c_{i_1}$ (i.e. the partial bijection that maps $c_{i_1}$ to $c_{1}$), and responds with $\pi^{-1}_1 \circ \varepsilon_1(\pi_1(x_1),y_1)=\pi^{-1}_1 \circ \varepsilon_1(c_1,y_1)$ for $y_1$. At the $j+1$th round, Universal plays some $c_{i_{j+1}}$ for $x_{j+1}$. If Universal has already played this, then Existential sets $\pi_{j+1}:=\pi_j$; otherwise Existential sets $\pi_{j+1}:=(c_{i_{j+1}} \mapsto c_{j+1}) \circ \pi_j $ (which also equals $(c_{i_{j+1}} \mapsto c_{j+1}) \uplus \pi_j$). In both cases she responds with 
\[ \pi^{-1}_{j+1} \circ \varepsilon_{j+1}(\pi_{j+1}(x_1),\pi_{j+1}(y_1),\ldots,\pi_{j+1}(x_{j+1})) \]
for $y_{j+1}$.
Since the strategy $\varepsilon$ is constant-conservative, no new constants are introduced through $\varepsilon$, and it follows from Lemma~\ref{lem:aut} that the strategy $\varepsilon'$ is winning.
\qed
\begin{remark}
Although the constant-conservative nature of Existential's play is used in the proof of $(ii \Rightarrow iii)$ above, it is only a truly vital component in the proof of $(iv \Rightarrow i)$. Imagine the play were not constant-conservative in that proof. Universal begins in the $\psi$-rel-game on $\mathcal{T}_\varphi(C_\omega)$ by playing $c_{i_1}$ for $x_1$, and Existential sets $\pi_1:=(c_{i_1},c_1)$. In the auxiliary $\psi$-rel-cc-game on $\mathcal{T}^{l^{l+2}}_\varphi(C_l)$, Existential now looks up what she would have played in her winning strategy if Universal had played $c_1$ for $x_1$. But, she might have played a response for $y_1$ that contains more than one constant. Clearly there is now the possibility to overload on constants in the auxiliary game.
\end{remark}

\paragraph{Illustration of the proof of Proposition~\ref{prop:cc} $(i \Rightarrow ii)$ by example}
Let $\varphi:=\forall x \forall z \exists y \ E(x,y) \wedge E(y,z)$ be as in Example~\ref{ex:main} and let $\psi :=$ 
\[
\begin{array}{ll}
\forall w_1 \exists w_2 \forall w_3 \exists w_4 \forall w_5 \exists w_6 & E(w_1,w_2) \wedge E(w_1,w_4) \wedge \\
& E(w_4,w_3) \wedge E(w_6,w_3). 
\end{array}
\]
Note that $w_5$ is essentially a dummy variable in $\psi$, but that $\psi$ (unlike $\varphi$) is in the correct normal form. Note also that $\models \varphi \rightarrow \psi$ (in fact, $\models \varphi \leftrightarrow \psi$).

The following is part of a game tree $\mathscr{G}_\varepsilon$ for the $\psi$-rel-game on $\mathcal{T}_\varphi(C_\omega)$ corresponding to a certain winning Existential strategy $\varepsilon$. Only the branches corresponding to Universal plays of the first three constants $c_1,c_2,c_3$ are depicted, and, even then, dashed arrows designate parts of the tree not expanded beyond their destination.

\[
\xymatrix{
 & & \bullet \ar[dl] \ar[d] \ar[dr] & \\
\forall w_1 & c_1 \ar@{-->}[d] & c_2 \ar[d] & c_3 \ar@{-->}[d]  \\
\exists w_2 & f(c_1,c_1) &*+[F]{f(c_2,c_1)} \ar[dl] \ar[d] \ar[dr]  & f(c_3,c_3)  \\
\forall w_3 & c_1 \ar@{-->}[d] & c_2 \ar[d] & c_3 \ar@{-->}[d]  \\
\exists w_4 & f(c_2,c_1) & f(c_2,c_2) \ar[dl] \ar[d] \ar[dr] & f(c_2,c_3)  \\
\forall w_5 & c_1 \ar[d] & c_2 \ar[d] & c_3 \ar[d]  \\
\exists w_6 & f(c_1,c_2) & f(c_1,c_2) & f(c_1,c_2) \\
}
\]
\medskip

It is easily seen that the strategy $\varepsilon$ is not constant-conservative, as attested by the boxed play of $f(c_2,c_1)$ for $w_2$. Below, we illustrate the technique for amending $\varepsilon$, so as to make it constant-conservative. At this node, the problem arises from Existential playing a term involving $c_1$, when Universal has not yet played $c_1$. Two branches beyond this node, corresponding to Universal plays of $c_1$ for $w_3$ and $w_5$, must be removed. And, in this node and any beyond, $c_1$ must be substituted by $c_2$ (the only constant thus far played by Universal). The tree so obtained is illustrated below, with the amended nodes highlighted.

\[
\xymatrix{
 & & \bullet \ar[dl] \ar[d] \ar[dr] & \\
\forall w_1 & c_1 \ar@{-->}[d] & c_2 \ar[d] & c_3 \ar@{-->}[d]  \\
\exists w_2 & f(c_1,c_1) & *+[F]{f(c_2,c_2)}  \ar[d] \ar[dr]  & f(c_3,c_3)  \\
\forall w_3 & & c_2 \ar[d] & c_3 \ar@{-->}[d]  \\
\exists w_4 & & f(c_2,c_2) \ar[d] \ar[dr] & f(c_2,c_3)  \\
\forall w_5 & & c_2 \ar[d] & c_3 \ar[d]  \\
\exists w_6 & & *+[F]{f(c_2,c_2)} & *+[F]{f(c_2,c_2)} \\
}
\]\medskip

\noindent It is now necessary to return the two branches corresponding to Universal plays of $c_1$ for $w_3$ and $w_5$.

\[
\xymatrix{
 & & \bullet \ar[dl] \ar[d] \ar[dr] & \\
\forall w_1 & c_1 \ar@{-->}[d] & c_2 \ar[d] & c_3 \ar@{-->}[d]  \\
\exists w_2 & f(c_1,c_1) & f(c_2,c_2)  \ar[dl] \ar[d] \ar[dr]  & f(c_3,c_3)  \\
\forall w_3 & c_1 \ar@{-->}[d] & c_2 \ar[d] & c_3 \ar@{-->}[d]  \\
\exists w_4 & f(c_2,c_1) & f(c_2,c_2) \ar[dl] \ar[d] \ar[dr] & f(c_2,c_3)  \\
\forall w_5 & c_1 \ar[d] & c_2 \ar[d] & c_3 \ar[d]  \\
\exists w_6 & f(c_2,c_2) & f(c_2,c_2) & f(c_2,c_2) \\
}
\]
\medskip

\noindent Note that $c_3$ has not been played on the path that now reads $c_2,f(c_2,c_2)$. We may therefore take the permutation that swaps $c_1$ and $c_3$ to replace the missing branch at $w_3$. Neither is $c_3$ played on the path $c_2,f(c_2,c_2),c_2,f(c_2,c_2)$, and we may take the same permutation to replace the missing branch at $w_5$.

\subsection{An Algorithm for Entailment}
\label{LHS:sec:alg}

Our decision procedure for the entailment problem makes use of the following fact, which may be proved by induction on $m$.
\begin{fact}
\label{fact2}
If $\varphi$ is a pH sentence of depth $k$, then $|T^m_\varphi(C_l)| \leq l^{(k+1)^m}$.
\end{fact}
\proof
Let $\tau:=|T^m_\varphi(C_l)|$ and $\tau':=|T^{m+1}_\varphi(C_l)|$. Clearly, $\tau'=1+\tau+\tau^2+\ldots+\tau^k \leq \tau^{k+1}$ (as $\tau\geq 2$) and the result follows.
\qed

\begin{thm}
The entailment problem for pH sentences is decidable in triple exponential time.
\end{thm}
\proof
Consider the input sentences $\varphi$ and $\psi$ of depth $k$ and $l$, respectively. By Theorem~\ref{thm:methodology} and Proposition~\ref{prop:cc}, it suffices to verify whether Existential has a winning strategy in the $\psi$-rel-cc-game on $\mathcal{T}^{l^{l+2}}_\varphi(C_l)$. The structure $\mathcal{T}^{l^{l+2}}_\varphi(C_l)$ is of size bounded by
\[ \zeta \ := \ (l+1) \uparrow (k+1) \uparrow (l) \uparrow (l+2),\]
where the $\uparrow$ denotes exponentiation (with precedence to the right).
We may search through all $2l$-tuples that could be played in the $\psi$-rel-cc-game on $\mathcal{T}^{l^{l+2}}_\varphi(C_l)$, in time $O(\zeta^{2l})$ to determine whether Existential has a winning strategy. Noting that
\[ \zeta^{2l} \ = \ \mathcal{O} ( (l+1) \uparrow (k+1) \uparrow (l) \uparrow 2l(l+2)), \]
the result follows.
\qed

\subsection{Undecidability of Entailment for Positive (equality-free) fo}
\label{LHS:sec:undecidable}

The \emph{entailment problem for positive fo} (\textsc{EPPFO}) is defined as follows.
\[
\begin{array}{ll}
\bullet & \mbox{Input: two sentences $\varphi$ and $\psi$ of positive (equality-} \\
& \mbox{free) fo.} \\
\bullet & \mbox{Question: does $\models \varphi \rightarrow \psi$?} 
\end{array}
\]
We consider also its dual problem, \textsc{Dual-EPPFO}.
\[
\begin{array}{ll}
\bullet & \mbox{Input: two sentences $\varphi$ and $\psi$ of positive (equality-} \\
& \mbox{free) fo.} \\
\bullet & \mbox{Question: is $\varphi \wedge \neg \psi$ satisfiable?} 
\end{array}
\]
These problems are clearly Turing equivalent ($\varphi \wedge \neg \psi$ is satisfiable iff it is not the case that $\neg \varphi \vee \psi$ is valid), and undecidability of the latter implies undecidability of the former.

We introduce one further problem, which may be seen as the satisfiability version of the (pure predicate) Classical Decision Problem, \textsc{Sat-CDP}.
\[
\begin{array}{ll}
\bullet & \mbox{Input: a sentence $\varphi$ of (equality-free) fo.} \\
\bullet & \mbox{Question: is $\varphi$ satisfiable?} 
\end{array}
\]
It is well-known that this problem is undecidable (see, e.g., \cite{CDP}). We are now in a position to prove the main result of this section.
\begin{thm}
The entailment problem for positive (equality-free) fo-logic, \\ \textsc{EPPFO}, is undecidable.
\end{thm}
\proof
By reduction from the \textsc{Sat-CDP} to the problem \textsc{Dual-EPPFO} defined above. Let $\varphi$ be some input to the \textsc{Sat-CDP}, containing relation symbols $R_1,\ldots,R_r$, of respective arities $a_1,\ldots,a_r$. We introduce $r$ new relation symbols $S_1,\ldots,S_r$, also of respective arities $a_1,\ldots,a_r$. We will now use these $S$-relations to axiomatise negation. Consider
\[ \theta_0 \ := \bigwedge_{i=1}^{r} \forall \tuple{x}_i  \ S_i(\tuple{x}_i) \leftrightarrow  \neg R_i(\tuple{x}_i) \]
\[ \theta_1 \ := \bigwedge_{i=1}^{r} \forall \tuple{x}_i  \ S_i(\tuple{x}_i) \vee R_i(\tuple{x}_i) \]
\[ \theta_2 \ := \bigwedge_{i=1}^{r} \forall \tuple{x}_i  \ \neg S_i(\tuple{x}_i) \vee \neg R_i(\tuple{x}_i), \]
where each $\tuple{x}_i$ is an $a_i$-tuple. Note that $\theta_0$ is logically equivalent to $\theta_1 \wedge \theta_2$. Now note that $\theta_2$ is logically equivalent to 
\[ \neg \bigvee_{i=1}^{r} \exists \tuple{x}_i  \ S_i(\tuple{x}_i) \wedge R_i(\tuple{x}_i),  \]
which we designate $\neg \psi$ (where $\psi$ is positive). Finally, derive $\varphi'$ from $\varphi$ by first propagating all negations to atomic level and then substituting any instances of negated relations $\neg R_i$ with $S_i$. It is easy to see that $\varphi$ is satisfiable iff $(\varphi' \wedge \theta_1) \wedge \neg \psi$ is satisfiable. Furthermore, $\varphi' \wedge \theta_1$ and $\psi$ are (equality-free) positive, and the result follows.
\qed

\section{Introducing Q-cores}
\label{sec:Q-cores}

\subsection{Canonical representatives and Core-ness}
\label{sec:core-ness}

A \emph{core} can be defined in various way, for example on finite structures one may say it is any structure all of whose endomorphisms are automorphisms. Consider the equivalence relation $\sim_{\mathrm{pp}}$ for finite structures induced by $\mathcal{A} \sim_{\mathrm{pp}} \mathcal{B}$ iff  $\mathrm{CSP}(\mathcal{A}) = \mathrm{CSP}(\mathcal{B})$ (\mbox{i.e.} $\mathcal{A}$ and $\mathcal{B}$ agree on all pp sentences). It is well-known that every member of each equivalence class of  $\sim_{\mathrm{pp}}$ contains, as an induced substructure, an isomorphic copy of the same core, which is (of course) also a member of that class. The core is thus uniquely minimal in its class with respect to both size and inclusion. Thus, for CSP and primitive positive logic the problem to find a canonical representative of the class induced by $\sim_{\mathrm{pp}}$ is straightforward (although still NP-hard!). Furthermore, each core $\mathcal{C}$ of size $n$ enjoys the property that there is a pp-formula $\phi(v_1,\ldots,v_n)$, so that $\mathcal{D}_\phi$ is an  isomorphic copy of that core, whose evaluation on $\mathcal{C}$ induces an isomorphism from $\mathcal{D}_\phi$ to $\mathcal{C}$ (we will paraphrase this by saying \emph{the constants are pp-definable in $\mathcal{C}$}). In particular, each element of $\mathcal{C}$ is individually pp-definable up to isomorphism.

What of a similar canonical representative for  $\sim_{\mathrm{pH}}$? We \textbf{might} try to call a structure $\mathcal{B}$ a ``Q-core'' if there is no pH-equivalent $\mathcal{A}$ of strictly smaller cardinality. We will discover that this ``Q-core'' would be a more cumbersome beast than its cousin the core; it need not be unique nor sit as an induced substructure of the templates in its class. However, in several cases we shall see in Section~\ref{sec:QcoresUseful} that its behaviour is reasonable and that -- like the core -- it can be very useful in delineating complexity classifications. 

We return to consider the following increasingly stronger fragments of fo logic:
\begin{enumerate}
\item primitive positive ($\csplogic$)
\item positive Horn, equality-free ($\qcsplogic$)\footnote{We specifically choose the equality-free version so that these four logics form a chain.}
\item positive equality-free fo (\mylogic); and,
\item positive fo (\posFO)
\end{enumerate}

The erratic behaviour of Q-cores sits in contrast not just to that of
cores, but also that of the \emph{$U$-$X$-cores} of~\cite{LICS2011},
which are the canonical representatives of the equivalence classes
associated with positive equality-free logic,  and were instrumental in deriving a full complexity
classification -- a tetrachotomy -- for its associated model-checking problem in \cite{LICS2011}. 
Like cores, they are unique 
and sit as induced substructures in all templates in their
class. Thus, primitive positive logic and positive equality-free logic
behave genially in comparison to their wilder cousin positive Horn. In
fact this manifests on the algebraic side also -- polymorphisms and
surjective hyper-endomorphisms 
are preserved under composition, while surjective polymorphisms are not. 

Continuing to add to our logics, in restoring equality, we might
arrive at positive logic. Two finite structures agree on all sentences
of positive logic iff they are isomorphic -- so here every finite
structure satisfies the ideal of ``core''. 
When computing a/the smallest substructure with the same behaviour with
respect to the four decreasingly weaker logics -- positive logic,
positive equality-free, positive Horn, and primitive positive --
we will obtain potentially structure decreasing in size. In the case
of positive equality-free and primitive positive logic, as pointed
out, these are unique up to isomorphism; and for the $U$-$X$-core
and the core, these will be induced substructures. A ``Q-core'' will
necessarily contain the core and be included in the $U$-$X$-core.  
This phenomenon is illustrated on Table~\ref{tab:different-cores} and
will serve as our running example.

\begin{table}[h]
  \centering
  \begin{tabular}[m]{|c|c|c|c|}
    \hline
    \posFO& \mylogic& \qcsplogic& \csplogic\\
    \hline
    $\mathcal{A}_4$&$\mathcal{A}_3$&$\mathcal{A}_2$&$\mathcal{A}_1$\\
    \begin{minipage}[c]{.25\textwidth}
      \centering
      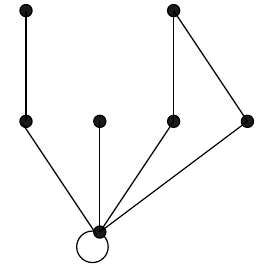
    \end{minipage}
    & 
    \begin{minipage}[c]{.2\textwidth}
      \centering
      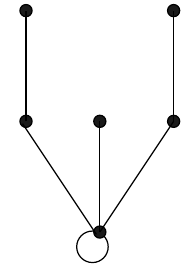
    \end{minipage}
    &
    \begin{minipage}[c]{.2\textwidth}
      \centering
      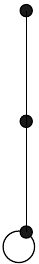
    \end{minipage}
    & 
    \begin{minipage}[c]{.15\textwidth}
            \centering
      \begin{picture}(0,0)%
\includegraphics{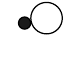}%
\end{picture}%
%
%
\setlength{\unitlength}{3108sp}%
\begingroup\makeatletter\ifx\SetFigFont\undefined%
\gdef\SetFigFont#1#2#3#4#5{%
  \reset@font\fontsize{#1}{#2pt}%
  \fontfamily{#3}\fontseries{#4}\fontshape{#5}%
  \selectfont}%
\fi\endgroup%
\begin{picture}(388,373)(4351,-5926)
\put(4366,-5911){\makebox(0,0)[lb]{\smash{{\SetFigFont{9}{10.8}{\rmdefault}{\mddefault}{\updefault}{\color[rgb]{0,0,0}0}%
}}}}
\end{picture}%

    \end{minipage}
    \\
    &&&\\[-8pt]
    \hline
    isomorphism & $U$-$X$-Core & Q-core & Core\\
    \hline
  \end{tabular}
  \caption{different notions of ``core'' (the circles represent self-loops).}
  \label{tab:different-cores}
\end{table}

\subsection{The case of QCSP}
\label{sec:qcsp}

In pp and pH logic, one normally considers
equalities to be permitted. From the perspective of computational
complexity of CSP and QCSP, this distinction is unimportant as
equalities may be propagated out by substitution. In the case of
pH and QCSP, though, equality does allow the distinction of
a trivial case that can not be recognised without it. The sentence
$\exists x \forall y \ x=y$ is true exactly on structures of size one (\mbox{cf.} Section~\ref{sec:LHS} and the ``degenerate'' cases). The structures $\mathcal{K}_1$ and $2 \mathcal{K}_1$, containing empty relations over one element and two elements, respectively, are therefore distinguishable in $\qcsplogiceq$, but not in $\qcsplogic$. Note that equalities can not be substituted out from $\posFO$, thus it is substantially stronger than $\mylogic$. In the previous parts of the paper, pH was generally assumed to contain equality (note that there is no $m$ so that ${\mathcal{K}_1}^m \surhom 2\cdot \mathcal{K}_1$). However, in this section, we will consider pH to be without equality, as it makes our chain of fours fragments of fo, from the previous section, increasing. For structures of size $>1$, expressibility in $\qcsplogiceq$ and $\qcsplogic$ coincide; thus to reconcile the different definitions of pH it is enough to limit ourselves to such structures. 


For pH, the correct concept to transfer winning strategies
is that of \emph{surjective homomorphism from a power}, something that we established already in Section~\ref{sec:RHS}.
Following our approach for the other logics, we now define a minimal
representative as follows.
\begin{definition} 
  A \emph{Q-core} $\mathcal{B}$ of a structure $\mathcal{A}$ is a minimal under inclusion
  substructure of $\mathcal{A}$ such that for every sentence $\varphi$
  in pH, $\mathcal{A}\models \varphi$ if and only if
  $\mathcal{B}\models \varphi$. 
\end{definition}
\begin{figure}[h]
  \centering
  \subfloat[$\mathcal{A}_2\times \mathcal{A}_2$.]{\label{fig:squareA2}
    \begin{minipage}[c]{.3\textwidth}
      \centering
      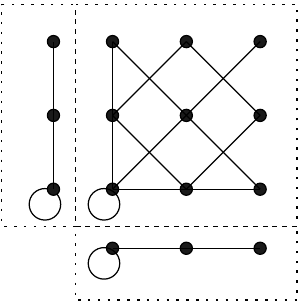
    \end{minipage}
}
  \qquad
  \subfloat[Homomorphism to $\mathcal{A}_3'$.]{\label{fig:surjhom:squareA2:A3}
    \begin{minipage}[c]{.3\textwidth}
      \centering
      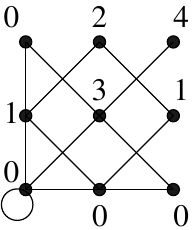
    \end{minipage}
}
  \qquad  
  \subfloat[$\mathcal{A}_3'$]{\label{fig:A3prime}
    \begin{minipage}[c]{.2\textwidth}
      \centering
      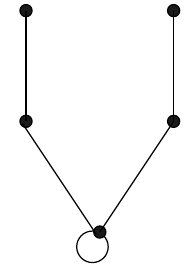
    \end{minipage}
}
  \qquad
  \caption{surjective homomorphism from a power.}
  \label{fig:squareA3}
\end{figure}

\begin{figure*}
  \centering
  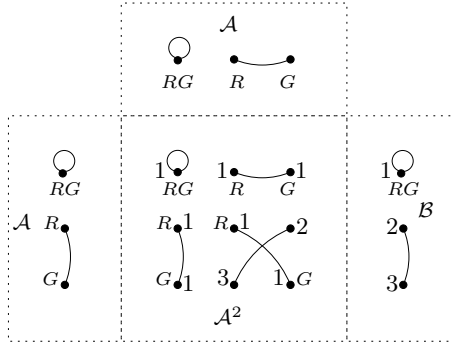  
  \caption{example of two distinct 3-element structures (signature, $E$ binary and two unary
    predicates $R$ and $G$) that are equivalent
    w.r.t. pH.}
  \label{fig:qcores}
\end{figure*}

\begin{ex}
  Consider $\mathcal{A}_3$ and $\mathcal{A}_2$ from
  Table~\ref{tab:different-cores}.
  We consider the subgraph $\mathcal{A}_3'$ of $\mathcal{A}_3$ as
  depicted on Figure~\ref{fig:A3prime}.
  The map $f(0):=0$, $f(1):=1$, $f(2):=2$, $f(3):=0$, $f(4):=0$
  is a surjective homomorphism from 
  $\mathcal{A}_3'$ to $\mathcal{A}_2$.
  The square of $\mathcal{A}_2$  is depicted on
  Figure~\ref{fig:squareA2};
  and, a surjective homomorphism from it to $\mathcal{A}_3'$ is
  depicted on Figure~\ref{fig:surjhom:squareA2:A3}. 
  Thus $\mathcal{A}_3'$ and $\mathcal{A}_2$ are equivalent
  w.r.t. pH.
  In a similar fashion but using a cube rather than a square, one can
  check that $\mathcal{A}_3$ and $\mathcal{A}_2$ are equivalent
  w.r.t. pH.
  One can also check that $\mathcal{A}_2$ is minimal and is therefore
  a Q-core of $\mathcal{A}_3$, and \textsl{a posteriori} of
  $\mathcal{A}_4$.
\end{ex}
The behaviour of the Q-core differs from its cousins the core and the
$U$-$X$-core.
\begin{prop}
A Q-core of a $3$-element structure $\mathcal{A}$ is not always an induced substructure of $\mathcal{A}$.
\end{prop}
\proof
  Consider the signature $\sigma:=\langle E,R,G\rangle$ involving a
  binary relation $E$ and two unary relations $R$ and $G$. Let
  $\mathcal{A}$ and $\mathcal{B}$ be structures with domain
  $\{1,2,3\}$ with the following relations. 
  \[ 
  \begin{array}{ccc}
    E^{\mathcal{A}}:=\{(1,1),(2,3),(3,2)\} & R^\mathcal{A}:=\{1,2\} & G^\mathcal{A}:=\{1,3\} \\
    E^{\mathcal{B}}:=\{(1,1),(2,3),(3,2)\} & R^\mathcal{B}:=\{1\} & G^\mathcal{B}:=\{1\}
  \end{array}
  \]
  Since $\mathcal{B}$ is a substructure of $\mathcal{A}$, we have $\mathcal{B} \surhom \mathcal{A}$.
  Conversely, the square of $\mathcal{A}^2$ contains an edge that has
  no vertex in the relation $R$ and $G$, which ensures that $\mathcal{A}^2 \surhom \mathcal{B}$
  (in Figure~\ref{fig:qcores} this surjective homomorphism is given explicitly).
  We can further check that no two-element structure $\mathcal{C}$, and \emph{a
  fortiori} no two-element substructure of $\mathcal{A}$, agrees with
  them on pH, and the result follows.
\qed
We still do not know whether the Q-core of a structure is unique (up to isomorphism). We will explore in the following section Q-cores over some special classes and show that this notion behaves well in
these cases. 

\subsection{The usefulness of Q-cores}
\label{sec:QcoresUseful}

 We term graphs \emph{reflexive} when any vertex has a self-loop; \emph{partially reflexive} (p.r.) to emphasise that
any vertex may or may not have a self-loop; and, \emph{irreflexive}
when they have none.
A \emph{p.r. tree} may contain self-loops but no larger cycle $\mathcal{C}_n$ for
$n\geq 3$. A \emph{p.r. forest}  is the disjoint union of p.r. trees.

Since \mbox{p.r.} forests  are closed under substructures, we can be assured that a Q-core of a
\mbox{p.r.} forest 
is a \mbox{p.r.} forest. It is clear from
inspection that the Q-core of \mbox{p.r.} forest 
is unique up to isomorphism, but we do not prove this as it does not
shed any light on the general situation. The doubting reader may
substitute ``a/ all'' for ``the'' in future references to Q-cores in
this section.

The complexity classifications of~\cite{QCSPforests} were largely
derived using the properties of equivalence \mbox{w.r.t.} pH.
This will be the central justification for the following propositions.

Let $\mathcal{K}^\star_i$ and $\mathcal{K}_i$ be the reflexive and irreflexive
$i$-cliques, respectively. Let $[n]:=\{1,\ldots,n\}$. For $i \in [n]$
and $\alpha \in \{0,1\}^n$, let $\alpha[i]$ be the $i$th entry of
$\alpha$. For $\alpha \in \{0,1\}^*$, 
let $\mathcal{P}_{\alpha}$ be
the path with domain $[n]$ and edge set $\{ (i,j) : |j-i|=1 \} \cup \{
(i,i) : \alpha[i]=1 \}$. 

For a tree $\mathcal{T}$ and vertex $v \in T$, let $\lambda_T(v)$ be the shortest distance in $\mathcal{T}$ from $v$ to a looped vertex (if $\mathcal{T}$ is irreflexive, then $\lambda_T(v)$ is always infinite). Let $\lambda_T$ be the maximum of $\{\lambda_T(v):v \in T\}$. A tree is \emph{loop-connected} if the self-loops induce a connected subtree. A tree $\mathcal{T}$ is \emph{quasi-loop-connected} if either 1.) it is irreflexive, or 2.) there exists a connected reflexive subtree $\mathcal{T}_0$ (chosen to be \textbf{maximal}) such that there is a walk of length $\lambda_T$ from every vertex of $\mathcal{T}$ to $T_0$. 

\subsubsection{Partially reflexive forests}

A \emph{majority} polymorphism, of a structure $\mathcal{A}$, is a homomorphism $f$ from $\mathcal{A}^3$ to $\mathcal{A}$ that satisfies, for all $x,y \in A$, $f(x,x,y)=f(x,y,x)=f(y,x,x)=x$.
It is not true that, if $\mathcal{H}$ is a \mbox{p.r.} forest, then either $\mathcal{H}$ admits a majority polymorphism, and QCSP$(\mathcal{H})$ is in NL, or QCSP$(\mathcal{H})$ is NP-hard. However, the notion of Q-core restores a clean delineation (the following proposition is a rephrased version of the main result from \cite{QCSPforests}).
\begin{prop}
Let $\mathcal{H}$ be a \mbox{p.r.} forest. Then either the Q-core of $\mathcal{H}$ admits a majority polymorphism, and QCSP$(\mathcal{H})$ is in NL, or QCSP$(\mathcal{H})$ is NP-hard.
\end{prop}
\proof
  We assume that graphs have at least one edge, for otherwise the Q-core is
  $\mathcal{K}_1$. (Recall that this assumes equality is forbidden from the language. If equality is present then the corresponding notion of Q-core for $m$ disjoint copies of $\mathcal{K}_1$ is: $\mathcal{K}_1$, if $m=1$; and $2\cdot \mathcal{K}_1$ otherwise.)
  Irreflexive forests are a special case of bipartite graphs, which are
  all equivalent \mbox{w.r.t.} pH, their Q-core being $\mathcal{K}_2$ when
  they have no isolated vertex (see
  example~\ref{ex:bip}) and $\mathcal{K}_2 \uplus \mathcal{K}_1$ otherwise.

  We assume from now on that graphs have at least one
  self-loop.
  The one vertex case is $\mathcal{K}_1^\star$. We assume larger graphs from now
  on. If the graph contains an isolated element then its Q-core is $\mathcal{K}_1 \uplus \mathcal{K}_1^\star$.
  Assume from now on that the graph does not have an isolated element.

  We deal with the disconnected case first.
  If the graph is reflexive, then its Q-core is $\mathcal{K}_1^\star \uplus\mathcal{K}_1^\star$.
  Otherwise, the graph is properly partially reflexive in the sense
  that it embeds both $\mathcal{K}_1^\star$ and $\mathcal{K}_1$. If the graph has an
  irreflexive component then its Q-core is $\mathcal{K}_2 \uplus\mathcal{K}_1^\star$.
  If the graph has no irreflexive component, then its Q-core is
  $\mathcal{K}_1^\star \uplus \mathcal{P}_{10^\lambda}$ where $\lambda$ is the longest walk
  from any vertex to a self-loop. The equivalence follows from
  analysing surjective homomorphism from suitable powers and requires
  a little work. By assumption there is a homomorphism from the
  graph to $\mathcal{K}_1^\star \uplus \mathcal{P}_{10^\lambda}$, mapping a connected
  component that has a vertex witnessing $\lambda$ to
  $\mathcal{P}_{10^\lambda}$ and all other connected components to
  $\mathcal{K}_1^\star$.
  Conversely, observe that the square of $\mathcal{K}_1^\star \uplus \mathcal{P}_{10^\lambda}$
  can be mapped surjectively to $\mathcal{K}_1^\star \uplus 3\cdot \mathcal{P}_{10^\lambda}$ and
  that any $\mathcal{P}_{10^\lambda}$ can be mapped surjectively to some
  $\mathcal{P}_{10^{\lambda'}}$, for $\lambda'< \lambda$.
  Thus, using a suitable homomorphic image of a sufficiently large
  power, we obtain for each vertex $x$ of the graph with associated
  parameter $\lambda'$ a copy of $\mathcal{P}_{10^{\lambda'}}$ which we may use
  to cover $x$.
  Minimality follows from the fact 
  that the Q-core must not satisfy $\forall x \exists
  y_1,\ldots,$ $\exists y_{\lambda-1} \ E(x,y_1) \wedge E(y_1,y_2) \wedge \ldots
  \wedge E(y_{\lambda-2},y_{\lambda-1}) \wedge E(y_{\lambda-1},y_{\lambda-1})$ 
  and must be disconnected.

We now follow the classification of \cite{QCSPforests}. If a \mbox{p.r.}
forest contains more than one \mbox{p.r.} tree, then the Q-core is
among those formed from the disjoint union of exactly two (including
the possibility of duplication) of $\mathcal{K}_1$, $\mathcal{K}^\star_1$, $\mathcal{P}_{10^\lambda}$,
$\mathcal{K}_2$. Each of these singularly admits a majority polymorphism, therefore so does any of their disjoint unions. 

We now move on to the connected case, \mbox{i.e.} it remains to consider \mbox{p.r.} trees $\mathcal{T}$. If $\mathcal{T}$ is
irreflexive, then its Q-core is $\mathcal{K}_2$ or $\mathcal{K}_1$, which admit
majority polymorphisms. If $\mathcal{T}$ is loop-connected, then it admits a
majority polymorphism \cite{QCSPforests}. If $\mathcal{T}$ is
quasi-loop-connected, then it is QCSP-equivalent to one of its
subtrees that is loop-connected \cite{QCSPforests} which will be its
Q-core, and admits majority. In all other cases QCSP$(\mathcal{T})$ is NP-hard, and $\mathcal{T}$ does not
admit majority \cite{QCSPforests}. 
\qed

\subsubsection{Irreflexive Pseudoforests}

A \emph{pseudotree} is a graph that involves at most one cycle. A \emph{pseudoforest} is the disjoint union of a collection of pseudotrees.
\begin{prop}
Let $\mathcal{H}$ be an irreflexive pseudoforest. Then either the Q-core of $\mathcal{H}$ admits a majority polymorphism, and QCSP$(\mathcal{H})$ is in NL, or QCSP$(\mathcal{H})$ is NP-hard.
\end{prop}
\proof
We follow the classification of \cite{CiE2006}. If $\mathcal{H}$ is bipartite, then its Q-core is either $\mathcal{K}_2$, $\mathcal{K}_1$, $\mathcal{K}_2 \uplus \mathcal{K}_1$ (see \cite{LICS2008}) and this admits a majority polymorphism. Otherwise its Q-core contains an odd cycle, which does not admit a majority polymorphism, and QCSP$(\mathcal{H})$ is NP-hard.
\qed

\subsection{The question of idempotency}

\begin{figure*}
  \centering
$
\xymatrix{
1  &
0  \\
1 &
1  \ar@{<->}[l] \ar@{<->}[u] \ar@{<->}[ul]  \ar@{<->}@(dl,dr) \\
\\
}
$ \hspace{1cm} $\xymatrix{\\ \surhom\\ \\}$ \hspace{1cm}
$
\xymatrix{
0  \ar@{<->}[d]  \\
1 \ar@{<->}@(dl,dr)  \\
}
$
  \caption{Surjective homomorphism from ${\mathcal{P}_{01}}^2$ to $\mathcal{P}_{01}$.}
  \label{fig:p012top01}
\end{figure*}
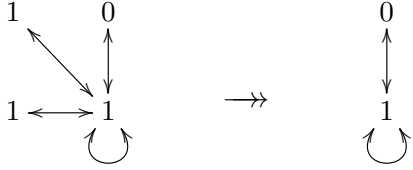

The observation was made in Section~\ref{sec:core-ness} that in a core $\mathcal{C}$ one can pp-define an isomorphic copy of the structure, which essentially renders the constants naming those elements to be pp-definable. We will now demonstrate that there is not always a representative of a class of $\sim_{\mathrm{pH}}$ in which this is possible, indeed we will give a class in which each member structure has elements that can not be pH-defined up to isomorphism. This class of $\sim_{\mathrm{pH}}$ has the unique Q-core $\mathcal{P}_{01}$ with vertices $\{0,1\}$ and edges $\{(0,1),(1,0),(1,1)\}$ (see Figure~\ref{fig:p012top01}). This is even a relatively well-behaved Q-core, uniquely of minimal cardinality and sitting as an induced substructure of everything in its class. 

A \emph{dominating} vertex in a graph $\mathcal{H}$ is some $x \in H$ so that for all $y \in H$ both $E(x,y)$ and $E(y,x)$ hold in $\mathcal{H}$ (this definition requires that $x$ be a self-loop). 
The members of $\mathcal{P}_{01}$'s equivalence class modulo $\sim_{\mathrm{pH}}$ are precisely those digraphs $\mathcal{H}$ with a dominating vertex and at least one vertex with no self-loop. To prove this in the forward direction we note that $\forall x \ E(x,x)$ and $\exists x \forall y \ E(x,y) \wedge E(y,x)$ are pH sentences. For the backward direction, observe that all such structures $\mathcal{H}$ have a surjective homomorphism to $\mathcal{P}_{01}$ (map some vertex with a non-loop to the non-loop and all other vertices to the loop) and a suitable power $r$ of $\mathcal{P}_{01}$ has a surjective homomorphism to $\mathcal{H}$ ($r:=|H|$ will do, see Figure~\ref{fig:p012top01} for an example with $r=2$).

Now, take any representative $\mathcal{H}$ of $\mathcal{P}_{01}$'s equivalence class modulo $\sim_{\mathrm{pH}}$ and some non-looped vertex $h_0 \in H$. We will argue that it is not possible to pH-define $h_0$ up to isomorphism, by showing that any pH-formula $\phi(x)$ that holds on $h_0$ in $\mathcal{H}$ actually holds on the dominating vertices of $\mathcal{H}$. Let $(\mathcal{H};h_0,h_1)$ be the expansion of $\mathcal{H}$ by constants $c_0$ and $c_1$ naming $h_0$ and some dominating vertex $h_1$, respectively. We argue that $(\mathcal{H};h_0,h_1)^2 \surhom (\mathcal{H};h_1,h_1)$, which tells us according to Theorem~\ref{theo:main:result} that if $\phi(x)$ holds on $h_0$ then it already holds on $h_1$! We illustrate this phenomenon specifically in the case $\mathcal{H}:=\mathcal{P}_{01}$ in Figure~\ref{fig:p012top01}, the generalisation to other representatives is clear.
Since there may not exist a representative of the  $\sim_{\mathrm{pH}}$ class in which the constants are definable, we must deduce that the reduction to the case of idempotent polymorphisms (in \mbox{e.g.} \cite{chen-2006}) is non-trivial. 

\section{Final Remarks}

\textbf{The model containment problem}. Two questions in particular arise from our discussion, and provide the most immediate challenge for further investigations.

We know that both the model containment problem for CSP and the CSP itself are NP-complete; indeed they are essentially the same problem. Given that the QCSP is Pspace-complete, it may be wondered what is the exact complexity of its associated model containment problem. It is far from clear that our algorithm is optimal; might the containment problem also be in Pspace, and, if so, might it be complete?

\textbf{The entailment problem}.
While $\models \varphi \rightarrow \psi$ is undecidable when both $\varphi$ and $\psi$ are positive fo, an analysis of our method yields that it is actually decidable for $\varphi$ pH and $\psi$ positive. This is because we may still build the canonical model of $\varphi$, and our game semantics hold in the presence of disjunction (these being essentially just Hintikka games; see \cite{Hodges}).

It is unclear how our method might be brought to bear on the question, for pH $\varphi$ and $\psi$, as to whether $\models_\mathrm{fin} \varphi \rightarrow \psi$ (\mbox{i.e.} the query containment problem for pH logic). If one could construct a finite canonical model $\mathcal{F}_\varphi$ for each $\varphi$, i.e. a finite model that still respects Theorem~\ref{thm:methodology} (Methodology II), one would have solved this. 

However, even for some simple sentences, we can demonstrate that there can be no finite canonical model. Consider $\varphi_1:= \forall x \exists y \ E(x,y)$, whose canonical models $\mathcal{T}_{\varphi_1}(C_1)$ and $\mathcal{T}_{\varphi_1}(C_\omega)$ are the infinite directed path ($\mathcal{DP}_\omega$) and $\omega$ disjoint copies of said path ($\mathcal{DP}_\omega \uplus \mathcal{DP}_\omega \uplus \ldots$), respectively. 

Suppose we had a finite model $\mathcal{F}_{\varphi_1}$ of size $d$ s.t., for all pH $\psi$, $\mathcal{F}_{\varphi_1} \models \psi$ iff $\models \varphi_1 \rightarrow \psi$. Since $\mathcal{F}_{\varphi_1} \models \varphi_1$, $\mathcal{F}_{\varphi_1}$ must contain (as a not-necessarily induced submodel) a directed cycle of length $e \leq d$ ($\mathcal{DC}_e$). It follows that the sentence $ \psi' :=$ 
\[
\begin{array}{ll}
\exists x_1,x_2,\ldots,x_{e-1},x_{e} & E(x_1,x_2) \wedge \ldots \wedge \\
& E(x_{e-1},x_{e}) \wedge E(x_{e},x_{1}) 
\end{array}
\]
is true on $\mathcal{F}_{\varphi_1}$. But $\varphi_1 \rightarrow \psi'$ is not logically valid, since $\mathcal{DC}_{e+1}$ is a model of the former but not the latter.

On the other hand, for some sentences we can produce finite canonical models. For $\varphi_2:= \forall x \exists y \ E(x,y) \wedge E(y,x)$, the finite canonical model $\mathcal{K}_2$ (or $\mathcal{K}_2 \uplus \mathcal{K}_2$) exists. That $\mathcal{K}_2$ is sufficient for this task follows from the fact that, for all models $\mathcal{A}$ of $\varphi_2$, there exists a constant $k_\mathcal{A}$ s.t. $(\mathcal{K}_2)^{k_\mathcal{A}} {\surhom} \mathcal{A}$, and therefore $\mathrm{QCSP}(\mathcal{K}_2) \subseteq \mathrm{QCSP}(\mathcal{A})$. Similarly, for $\varphi_3:=$
\[
\begin{array}{ll}
\forall x \exists y \exists z & E(x,y) \wedge E(y,x) \wedge E(y,z) \wedge \\
& E(z,y) \wedge E(z,x) \wedge E(x,z), 
\end{array}
\]
the canonical model $\mathcal{K}_3 \uplus \mathcal{K}_3$ exists. In the latter case $\mathcal{K}_3$ will not do: consider $\psi'':=$ 
\[ \forall x \forall y \exists w\exists z \ E(x,y) \wedge E(y,w) \wedge E(w,z) \wedge E(z,y); \]
$\varphi_3 \rightarrow \psi''$ is not logically valid, as $\mathcal{K}_3 \uplus \mathcal{K}_3$ models the former but not the latter, but $\mathcal{K}_3 \models \psi''$. 

This problem has been registered at \cite{FMT-open-problem-garden}.

\textbf{Q-cores}.
There are two outstanding questions here. Firstly, is the Q-core of a finite structure unique up to isomorphism (when one considers non-induced substructure). Secondly, for every finite $\mathcal{A}$ does there exists a finite $\mathcal{B}$ so that QCSP$(\mathcal{A})$ and QCSP$(\mathcal{B})$ are polynomial-time equivalent and the constants are (all-at-once) pH-definable in $\mathcal{B}$ (up to isomorphism). We know this is false with ``polynomial-time equivalent'' replace by ``equal'', but this indirect method may yet salvage the legitimacy to assume we can deal with idempotent polymorphisms alone.

\section*{Acknowledgements}

We are very grateful to Arnaud Durand for supplying the undecidability proof of the entailment problem for positive (equality-free) fo-logic. We are also grateful to an anonymous referee from the conference version for directing us to the paper of Keisler. Finally, we are grateful to a number of referees of the journal version for their corrections and patience.

\bibliographystyle{acm}

\end{document}